\documentclass[journal,twoside]{IEEEtran}
\usepackage{cite}

\ifCLASSINFOpdf
\else
   \usepackage[dvips]{graphicx}
\fi
\usepackage[cmex10]{amsmath}
\usepackage{amssymb,amsthm}
\usepackage{balance}

\newtheorem{definition}{Definition} 
\newtheorem{example}{Example} 
\newtheorem{remark}{Remark} 
\newtheorem{proposition}{Proposition} 
\newtheorem{lemma}{Lemma} 

\begin{document}
%
\title{Iterative LMMSE Channel Estimation, Multiuser Detection, and Decoding via Spatial Coupling}
%
%
%

\author{Keigo~Takeuchi,~\IEEEmembership{Member,~IEEE}
\thanks{
The work of K.~Takeuchi was in part 
supported by the Grant-in-Aid for Young Scientists (A) 
(MEXT/JSPS KAKENHI Grant Number 26709029), Japan.
}
\thanks{K.~Takeuchi is with the Department of Communication 
Engineering and Informatics, the University of Electro-Communications, 
Tokyo 182-8585, Japan (e-mail: ktakeuchi@uec.ac.jp).}
}

%
%

\markboth{IEEE transactions on information theory,~Vol.~, No.~, 2014}%
{Takeuchi \MakeLowercase{\textit{et al.}}:Iterative LMMSE Channel Estimation, Multiuser Detection, and Decoding via Spatial Coupling}
%

\IEEEpubid{0000--0000/00\$00.00~\copyright~2014 IEEE}


\maketitle

\begin{abstract}
Spatial coupling is utilized to improve the performance of iterative channel 
estimation, multiuser detection, and decoding for multiple-input 
multiple-input (MIMO) bit-interleaved coded modulation (BICM). 
Coupling is applied to both coding and BICM---the encoder uses a 
protograph-based spatially-coupled low-density parity-check (SC LDPC) code. 
Spatially and temporally coupled (STC) BICM is proposed to enable iterative 
channel estimation via coupling. Linear minimum mean-squared error (LMMSE) 
estimation is applied for both channel estimation and detection to reduce the 
complexity. Tractable density evolution (DE) equations are derived to analyze 
the convergence property of iterative receivers in the large-system limit, 
via a tool developed in statistical physics---replica method. The DE analysis 
implies that the STC BICM can improve the performance of iterative channel 
estimation especially for higher-order modulation. Numerical simulations show 
that the STC BICM can provide a significant gain of the performance at high 
signal-to-noise ratios for $64$ quadrature amplitude modulation (QAM), 
as well as an improvement in the decoding threshold, 
compared to conventional BICM.  
\end{abstract}

\begin{IEEEkeywords}
Spatial coupling, multiple-input multiple-output (MIMO) systems, 
iterative channel estimation, replica method, density evolution. 
\end{IEEEkeywords}

%

\section{Introduction}
\IEEEPARstart{H}{igh} spectral efficiency is required in modern wireless 
communications. Multiple-input multiple-output (MIMO) 
transmission~\cite{Foschini98,Telatar99} has been used to achieve 
high spectral efficiency.  
The channel capacity of MIMO systems grows in proportion to the minimum of 
the numbers of transmit and receive antennas~\cite{Telatar99}. 
It is an important issue to construct low-complexity receivers that 
realize this potential.  

Since the invention of turbo codes~\cite{Berrou96}, iterative multiuser 
detection and decoding (MUDD)~\cite{Alexander98,Reed98} has been 
proposed as a promising scheme for solving that issue. When 
bit-interleaved coded modulation (BICM) is used, iterative MUDD can be 
formulated in a unified framework based on belief propagation 
(BP)~\cite{Pearl88,Boutros02}. The optimal symbol-wise  
maximum-a-posteriori (MAP) detection results in high complexity, so that 
the linear minimum mean-squared error (LMMSE) detection has been used  
instead in \cite{Wang99}. The iterative LMMSE MUDD can achieve excellent 
decoding performance in spite of its low complexity. 

Iterative MUDD based on BICM has been extended to iterative channel estimation 
and MUDD (CE-MUDD)~\cite{Alexander00,Gamal00,Kobayashi01,Lampe02,Valenti01,Loncar04,Rossi08}. Since the optimal nonlinear channel estimator has impractical 
complexity, the LMMSE channel estimator~\cite{Valenti01,Loncar04} has been 
used instead, as considered in iterative MUDD. 
The iterative LMMSE CE-MUDD can provide a significant reduction of the 
overhead for training. 

In this paper, we improve the performance of iterative CE-MUDD 
for MIMO BICM systems via spatial coupling. 
Spatial coupling\footnote{
Coupling is actually made in the temporal domain for coding. Thus, 
spatial coupling for coding should be regarded as technical terminology. 
On the other hand, coupling we introduce for BICM is in the spatial domain 
for MUDD, whereas it is in the temporal domain for channel estimation.} 
was proposed as a sophisticated technique for improving the BP performance of 
low-density parity-check (LDPC) codes up to the MAP 
performance~\cite{Kudekar11,Lentmaier102}. A spatially coupled (SC) LDPC 
code is constructed as a chain of multiple LDPC codes. Both ends of the chain 
are terminated so that the bits at both ends are decoded successfully. 
The reliable information at both ends can propagate toward the center of the 
chain without error propagation, when the code length in each section of 
the chain is sufficiently long. The rate loss due to the termination is 
negligible when the chain is sufficiently long. As a result, the BP 
decoder for the SC LDPC code can achieve the MAP performance of the 
underlying LDPC code. This improvement in performance via spatial coupling is 
{\em universal}~\cite{Hassani12,Takeuchi122,Takeuchi142,Yedla12,Schlegel132,Khatib14}, and spatial coupling has been applied to many other 
problems~\cite{Takeuchi142,Takeuchi112,Takeuchi131,Schlegel131,Krzakala12,Donoho13,Kasai11,Yedla11,Aref13}. 
As one more application, spatial coupling is used to improve 
the performance of iterative CE-MUDD for MIMO systems with no 
channel state information (CSI).  

We apply coupling to both coding and BICM---we use a protograph-based 
SC LDPC code~\cite{Lentmaier102} in coding and a coupled interleaver in BICM. 
SC LDPC coding was applied to iterative MUDD for MIMO systems~\cite{Zhang12}. 
On the other hand, coupled interleavers were proposed to improve the 
performance of iterative MUDD for LDPC-coded MIMO systems with perfect CSI 
at the receiver~\cite{Takeuchi132} and of iterative channel estimation and 
decoding for single-antenna systems~\cite{Horio15}. In this paper, 
we propose spatially and temporally coupled (STC) BICM to improve 
the performance of iterative CE-MUDD. 

\IEEEpubidadjcol

The purpose of this paper is to investigate whether coupling should be 
introduced for coding or BICM. In previous works, spatial coupling was 
applied to either coding or BICM. However, to the best of author's knowledge, 
there are no comparisons between the two SC systems. In this paper, 
we introduce coupling for both coding and BICM, and elucidate whether coupling 
should be used in coding or BICM. 

Density evolution (DE) is used to analyze the performance of iterative CE-MUDD 
for MIMO systems with SC LDPC coding and STC BICM. DE is a powerful method for 
analyzing the convergence property of the BP decoder for LDPC codes or turbo 
codes when the code length tends to 
infinity~\cite{Richardson01,Chung01,Gamal01,Brink01}. The method has 
been used to analyze iterative MUDD~\cite{Boutros02,Caire04}, iterative 
channel estimation~\cite{Lee08,Takeuchi133} for single-input single-output 
systems, and iterative CE-MUDD~\cite{Vehkaperae09} for code-division 
multiple-access (CDMA) systems. We extend the DE analysis to the case of 
iterative CE-MUDD for MIMO systems with SC LDPC coding and STC BICM. 

The DE analysis for iterative MUDD cannot yield {\em analytical} DE 
equations for describing the convergence property as long as the system 
size is finite. Consequently, Monte Carlo simulations are required for solving 
the DE equations. In order to circumvent this difficulty, the large-system 
limit has been considered~\cite{Caire04,Vehkaperae09}, in which the numbers of 
transmit and receive antennas tend to infinity at the same rate. 
The large-system analysis results in analytical DE equations, which can 
provide a good approximation for small MIMO systems~\cite{Takeuchi134}. 

In this paper, the replica method is used for the large-system analysis. 
The method was originally developed in statistical 
physics~\cite{Mezard87,Nishimori01}, and introduced in the field of 
communications by Tanaka~\cite{Tanaka02}. Then, the replica method has been 
used for the large-system analysis of MIMO systems with perfect 
CSI~\cite{Moustakas03,Mueller04,Guo05,Wen07,Takeuchi08}. 
MIMO systems with no CSI have been analyzed in \cite{Takeuchi121,Takeuchi134}.  
In this paper, the large-system analysis in \cite{Takeuchi134} is applied 
to the case of MIMO systems with STC BICM. We note that the replica 
method is based on several non-rigorous assumptions, although it is believed 
in the literature that the replica method provides correct results.  

The contribution of this paper is to present an answer to the main 
question---whether coupling should be introduced for coding or BICM? 
We will show that STC BICM hardly improves the performance of 
iterative CE-MUDD for quadrature phase shift keying (QPSK), when SC LDPC coding 
is used. This result means that, for QPSK, it is sufficient to introduce 
coupling only in coding. For higher-order modulation, STC BICM can improve the 
performance of iterative CE-MUDD even when SC LDPC coding is used. 
Thus, we claim that coupling should be introduced for BICM or for both coding 
and BICM if MIMO systems with higher-order modulation are used. 

The remainder of this paper is organized as follows: After summarizing the 
notation used in this paper, In Section~\ref{section2} we present 
MIMO systems with SC LDPC coding and STC BICM. 
In Section~\ref{section3} we derive an iterative LMMSE CE-MUDD algorithm 
based on approximate BP. In Section~\ref{section4} the DE analysis of the 
iterative CE-MUDD is performed with the replica method. After presenting 
comparisons between conventional BICM and STC BICM in 
Section~\ref{section5}, we conclude this paper in Section~\ref{section6}. 

Throughout this paper, $\boldsymbol{A}^{\mathrm{T}}$ and 
$\boldsymbol{A}^{\mathrm{H}}$ denote the transpose and conjugate transpose of 
a matrix $\boldsymbol{A}$, respectively. The complex conjugate of a complex 
number $z$ is denoted by $z^{*}$. We write a proper complex Gaussian 
distribution with mean $\boldsymbol{m}$ and covariance $\boldsymbol{\Sigma}$ 
as $\mathcal{CN}(\boldsymbol{m},\boldsymbol{\Sigma})$. For integers 
$i$ and $j$ ($i<j$), $[i:j)$ represents the set of integers 
$\{i,i+1,\ldots,j-1\}$. The set $[i:j]=\{i,\ldots,j\}$ is defined in the 
same manner. The notation $f(x)\propto g(x)$ implies that there is a positive 
constant $A>0$ such that $f(x)=Ag(x)$ for any $x$. 

\section{System Model} \label{section2} 
\subsection{MIMO System} 
We consider point-to-point MIMO systems with $K$ transmit antennas and $N$ 
receive antennas. For simplicity, block-fading channels with coherence 
time~$T$ are assumed---the channel matrix is fixed during $T$ symbol periods, 
and changes independently at the beginning of the next fading block. 
This channel model may be regarded as a mathematical model for time-division 
multiple-access (TDMA) or frequency-hopping systems~\cite{Marzetta99}. 
Furthermore, we assume independent Rayleigh fading. The channel matrix 
$\boldsymbol{H}(b)\in\mathbb{C}^{N\times K}$ in each fading block~$b$ has 
independent circularly symmetric complex Gaussian (CSCG) random elements with 
variance $1/K$. We note that the latter 
assumption is an idealized assumption to simplify the large-system analysis. 
There may be dependencies between the elements in practice. Under these 
assumptions, the received vector $\boldsymbol{y}_{t}(b)\in\mathbb{C}^{N}$ 
in symbol period~$t\in[1:T]$ within fading block~$b$ is given by 
\begin{equation} \label{MIMO} 
\boldsymbol{y}_{t}(b) 
= \boldsymbol{H}(b)\boldsymbol{x}_{t}(b) + \boldsymbol{w}_{t}(b). 
\end{equation}
In (\ref{MIMO}), $\boldsymbol{x}_{t}(b)=(x_{1,t}(b),\ldots,x_{K,t}(b))^{\mathrm{T}}$  
denotes the transmitted vector in symbol period~$t$ within fading block~$b$. 
Furthermore, $\boldsymbol{w}_{t}(b)\sim\mathcal{CN}(\boldsymbol{0},
N_{0}\boldsymbol{I}_{N})$ is independent additive white Gaussian noise (AWGN) 
vectors with covariance $N_{0}\boldsymbol{I}_{N}$ in the same symbol period. 
The channel matrices are unknown to the receiver in advance, whereas 
all statistical properties are assumed to be known. 

The MIMO channel~(\ref{MIMO}) can be expressed in matrix form as 
\begin{equation} \label{matrix_MIMO}
\boldsymbol{Y}(b) = \boldsymbol{H}(b)\boldsymbol{X}(b) 
+ \boldsymbol{W}(b), 
\end{equation}
with the received matrix 
$\boldsymbol{Y}(b)=(\boldsymbol{y}_{1}(b),\ldots,\boldsymbol{y}_{T}(b))$, 
the transmitted matrix 
$\boldsymbol{X}(b)=(\boldsymbol{x}_{1}(b),\ldots,\boldsymbol{x}_{T}(b))$, 
and the AWGN matrix 
$\boldsymbol{W}(b)=(\boldsymbol{w}_{1}(b),\ldots,\boldsymbol{w}_{T}(b))$. 

\begin{figure}[t]
\begin{center}
\includegraphics[width=\hsize]{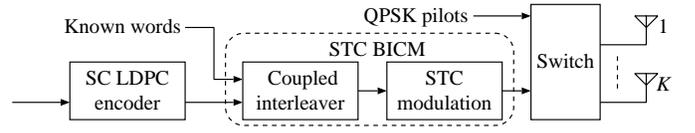}
\end{center}
\caption{
SC-LDPC-coded MIMO STC BICM. 
}
\label{fig0}
\end{figure}

We consider a pilot-assisted MIMO STC BICM scheme, shown in Fig.~\ref{fig0}. 
In order to estimate the channel matrices, the first $T_{\mathrm{tr}}$ symbol 
periods in each fading block are assigned to transmission of QPSK pilot 
symbols with unit power $|x_{k,t}(b)|^{2}=1$, so that 
the matrix $\boldsymbol{X}_{[1:T_{\mathrm{tr}}]}(b)=(\boldsymbol{x}_{1}(b),\ldots,
\boldsymbol{x}_{T_{\mathrm{tr}}}(b))$ in the training period $t\in[1:T_{\mathrm{tr}}]$ 
is assumed to be known for all $b$. The remaining symbol periods are assigned 
to transmission of data symbols. In the following sections, 
we present the details of the SC LDPC encoder and the STC BICM. 

\subsection{Spatially Coupled LDPC Codes}
\begin{figure}[t]
\begin{center}
\includegraphics[width=\hsize]{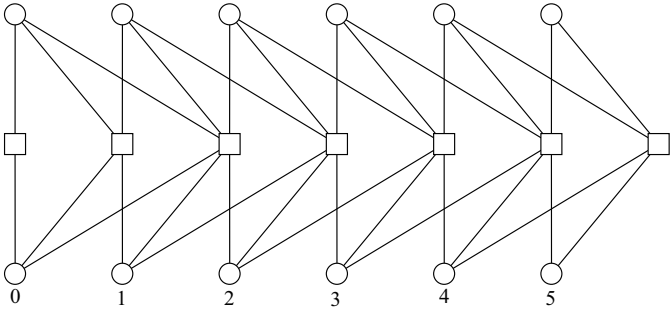}
\end{center}
\caption{
Protograph for $(3, 6, 6)$ ensemble of SC LDPC codes. The variable and check 
nodes are represented by circles and boxes, respectively. 
}
\label{fig1}
\end{figure}

We shall review the ensemble of protograph-based SC LDPC codes with the degree 
of variable nodes~$d_{\mathrm{v}}$, the degree of check nodes~$d_{\mathrm{c}}$, and 
the number of sections~$L$. Efficient termination~\cite{Tazoe12} is used 
to reduce the computational complexity in encoding. We refer to this ensemble 
as $(d_{\mathrm{v}}, d_{\mathrm{c}}, L)$ ensemble. 
See Fig.~\ref{fig1} for an example of the protograph that represents the 
$(3, 6, 6)$ ensemble, which is defined via the $(L+1)\times
(d_{\mathrm{c}}/d_{\mathrm{v}})L$ base matrix 
\begin{equation} \label{base_matrix}
\boldsymbol{B}
= \left(
\begin{array}{cccccccccccccc}
1 & 1 & 0 & 0 & 0 & 0 & 0 & 0 & 0 & 0 & 0 & 0 \\
1 & 1 & 1 & 1 & 0 & 0 & 0 & 0 & 0 & 0 & 0 & 0  \\ 
1 & 1 & 1 & 1 & 1 & 1 & 0 & 0 & 0 & 0 & 0 & 0  \\
0 & 0 & 1 & 1 & 1 & 1 & 1 & 1 & 0 & 0 & 0 & 0  \\
0 & 0 & 0 & 0 & 1 & 1 & 1 & 1 & 1 & 1 & 0 & 0  \\
0 & 0 & 0 & 0 & 0 & 0 & 1 & 1 & 1 & 1 & 1 & 1  \\
0 & 0 & 0 & 0 & 0 & 0 & 0 & 0 & 1 & 1 & 1 & 1  \\
\end{array}
\right).
\end{equation} 

The base matrix is obtained by removing the last $(d_{\mathrm{v}}-2)$ rows of 
the conventional $(L+d_{\mathrm{v}}-1)\times(d_{\mathrm{c}}/d_{\mathrm{v}})L$ base 
matrix proposed in \cite{Lentmaier102}. 
Removing the last rows results in performance degradation for finite-length 
codes, whereas the decoding threshold does not change~\cite{Tazoe12}. 
Roughly speaking, only reliable information at the left end propagates to 
the right side of the protograph when the signal-to-noise ratio (SNR) is 
close to the decoding threshold. The only reason why we use this ensemble 
is to reduce the encoding complexity in numerical simulations. It is 
straightforward to replace our results by those for the conventional ensemble 
without efficient termination~\cite{Lentmaier102} if another efficient method 
for encoding is available in the future.  

A parity-check matrix is generated from the base matrix~(\ref{base_matrix}) 
as follows: We replace all ones in the base matrix with 
$(d_{\mathrm{v}}/d_{\mathrm{c}})M\times (d_{\mathrm{v}}/d_{\mathrm{c}})M$ 
random permutation matrices independently. On the other hand, all zeros are 
replaced by all-zero matrices with the same size. The obtained 
$(L+1)(d_{\mathrm{v}}/d_{\mathrm{c}})M\times LM$
parity-check matrix corresponds to an SC factor graph with $L$ sections. 
The design rate $r$ of the obtained SC LDPC code is given by 
\begin{equation} \label{design_rate} 
r = 1 - \frac{d_{\mathrm{v}}}{d_{\mathrm{c}}} 
- \frac{d_{\mathrm{v}}}{d_{\mathrm{c}}L}, 
\end{equation}
which tends to the design rate~$1-d_{\mathrm{v}}/d_{\mathrm{c}}$ of the underlying 
LDPC code as $L\to\infty$. 

In section~$l\in\{0,\ldots,L-1\}$ the graph has $M$ variable nodes, which are 
called a codeword in the section. Thus, the overall codeword of an SC LDPC 
code consists of $L$ codewords.

\subsection{Spatially and Temporally Coupled BICM} \label{sec2B}
\subsubsection{Coupled Interleaver}
As the first part of the STC BICM, we define coupled interleavers, 
which is slightly different from those proposed in \cite{Takeuchi132,Horio15}. 
Before presenting coupled interleavers, we shall introduce the ensemble 
of random interleavers. An interleaver of length~$M$ is a permutation 
of $\mathcal{M}=\{0,1,\ldots,M-1\}$. The ensemble of random interleavers 
consists of all possible permutations of $\mathcal{M}$. Each interleaver is 
picked up from the ensemble uniformly and randomly. It is known that 
performance for individual interleavers converges almost surely to its average 
over the ensemble when the length~$M$ tends to 
infinity~\cite{Boutros02,Caire04}.  
Thus, it is sufficient to investigate the average performance over the 
ensemble, instead of performance for individual interleavers. 

Let $L+W$ and $W$ denote the number of sections and coupling width, 
respectively. A coupled interleaver is constructed as follows: 
We first generate $L+W$ independent interleavers of length~$M$, and refer to  
the $l$th interleaver as section~$l$ at the input side 
for $l\in\mathcal{L}=\{-W,-W+1,\ldots,L-1\}$. Each section is divided into 
$2W+1$ subsections with size $\tilde{M}=M/(2W+1)$. 
Bit~$\mu\in\tilde{\mathcal{M}}=\{0,\ldots,\tilde{M}-1\}$ in 
subsection~$w\in\mathcal{W}=\{-W,-W+1,\ldots,W\}$ corresponds to the bit 
$m=\mu+(w+W)\tilde{M}\in\mathcal{M}$ in the same section.  
Equivalently, bit~$m\in\mathcal{M}$ in each section corresponds to the bit 
$\mu(m)=m-(w(m)+W)\tilde{M}$ in the subsection 
$w(m)=\lfloor m/\tilde{M}\rfloor-W\in\mathcal{W}$. 
Subsection~$w\in\mathcal{W}$ in section~$l\in\mathcal{L}$ is {\em coupled} 
with subsection~$-w$ in section~$l+w$ at the output side for $w\in\mathcal{W}$. 
If $l+w\notin\mathcal{L}$, it is coupled with subsection~$w$ in section $l$. 
The $(M,L,W)$ interleaver is formally defined as follows: 

\begin{figure}[t]
\begin{center}
\includegraphics[width=\hsize]{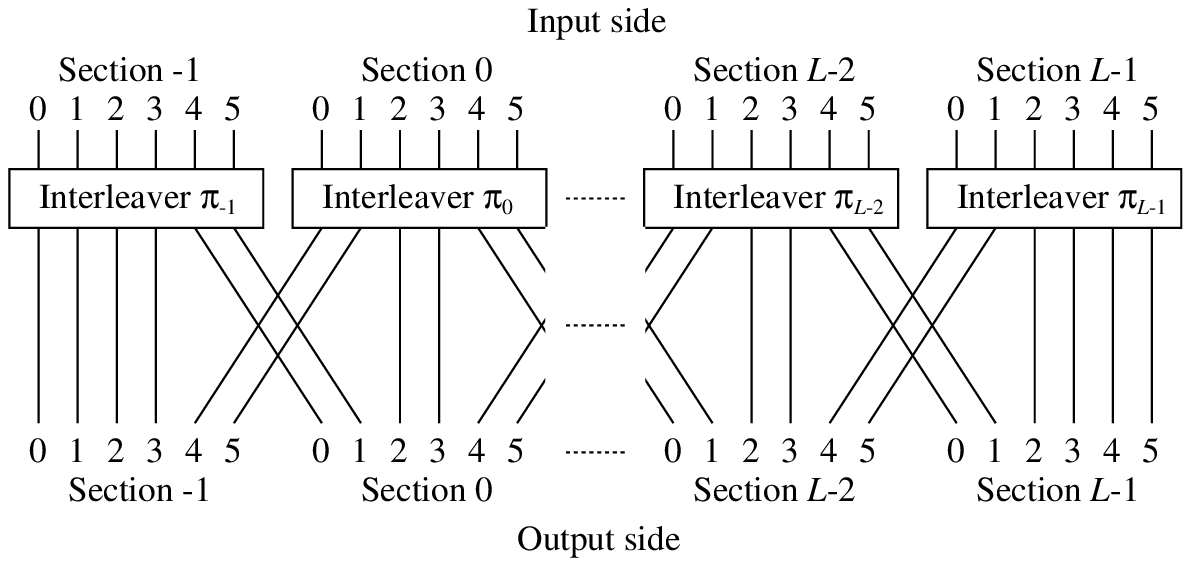}
\end{center}
\caption{
Coupled interleaver for $M=6$ and $W=1$. 
}
\label{fig2}
\end{figure}

\begin{definition}[$(M,L,W)$ Interleaver]
A coupled interleaver $\pi$ is a bijection from 
$\mathcal{M}\times\mathcal{L}$ onto 
$\tilde{\mathcal{M}}\times\mathcal{L}\times\mathcal{W}$. Picked up $L+W$ 
interleavers $\{\pi_{l}:l\in\mathcal{L}\}$ of length~$M$ from the ensemble of 
random interleavers uniformly and randomly. Let 
$\mu(\tilde{m})\in\tilde{\mathcal{M}}$ and $w(\tilde{m})\in\mathcal{W}$ 
denote the bit and subsection that correspond to the output bit 
$\tilde{m}=\pi_{l}(m)\in\mathcal{M}$. 
For $(m,l)\in\mathcal{M}\times\mathcal{L}$, 
the coupled interleaver $\pi(m,l)$ is given by  
\begin{equation} \label{coupled_interleaver} 
\pi(m,l) 
= \left\{
\begin{array}{cl}
(\mu(\tilde{m}),l+w(\tilde{m}),-w(\tilde{m})) 
& \hbox{for $l+w(\tilde{m})\in\mathcal{L}$} \\ 
(\mu(\tilde{m}),l,w(\tilde{m})) 
& \hbox{for $l+w(\tilde{m})\notin\mathcal{L}$,}
\end{array}
\right.
\end{equation}
with $\tilde{m}=\pi_{l}(m)\in\mathcal{M}$. 
\end{definition}

See Fig.~\ref{fig2} for the $(6,L,1)$ interleaver. 
We note that the $(M,L,W)$-interleaver reduces to conventional random 
interleavers for $W=0$. 

Known words are sent in sections~$l$ for $l<0$ in order to start up iterative 
CE-MUDD via coupling. On the other hand, the codeword in section~$l$ of the SC 
LDPC code is transmitted in section~$l$ of the coupled interleaver for the 
remaining sections $l\in\{0,\ldots,L-1\}$. Note that no known words are 
transmitted in the opposite end, because of the termination 
structure~(\ref{base_matrix}) of SC LDPC codes.  

\subsubsection{Spatially and Temporally Coupled Modulation}
We next present the mapping method from the interleaved bit sequences to 
data matrices. Each data matrix is generated so as to satisfy two conditions. 
One condition is that each data symbol originates from a codeword in the 
same section. The other condition is that data symbols generated from 
different codewords are included in each $K\times (T-T_{\mathrm{tr}})$ data 
matrix with spatially and temporally uniform probability. The terminology 
``STC'' BICM\footnote{
One should not confuse STC BICM with space-time coded BICM. 
} originates from the latter condition. The former condition  
simplifies the DE analysis. See \cite{Takeuchi141} for a relaxation of 
the former condition. 

We consider a constellation $\mathcal{C}\subset\mathbb{C}$ with 
modulation degree $Q=\log_{2}|\mathcal{C}|\in\mathbb{N}$. 
Our approach for satisfying the two conditions is as follows: 
$Q$ bits in each subsection are mapped to one data symbol in 
order to satisfy the former condition. If a data symbol originates  
from the bits in subsection~$w\in\mathcal{W}$ (respectively (resp.\ $W$)), 
the next symbol is generated from the bits in the next 
subsection~$w+1$ (resp.\ $-W$). Assume that the number $K$ of transmit 
antennas is a multiple of the number $2W+1$ of subsections, or that $K$ 
tends to infinity with $W$ fixed. When the data symbols are transmitted 
with $K$ transmit antennas after the generation of every $K$ data symbols, 
the data symbols in the same symbol period originate from different codewords 
with equal frequency. 

In order to realize uniformity in the temporal domain, 
we propose an assignment with a cyclic shift. 
In the proposed assignment, the first $K$ data symbols are transmitted from 
the transmit antennas~$k$ in the order~$k=1,\ldots,K$. The next $K$ symbols 
are assigned to the transmit antennas~$k$ in the 
order~$k=2,\ldots,K,1$, which is a cyclic shift of the assignment in the 
preceding symbol period. Such assignments with the cyclic shift are repeated 
for all symbol periods. 
Assume that $K$ and $T-T_{\mathrm{tr}}$ are a multiple of $2W+1$, or 
that $K$ and $T-T_{\mathrm{tr}}$ tend to infinity at the same rate. Then, 
each data stream contains data symbols originating from different codewords 
with equal frequency.  

If assignments with no cyclic shift were used, each transmit antenna would 
continue to emit the data symbols originating from the same codeword when $K$ 
is a multiple of $2W+1$. This implies that the channel estimator cannot utilize 
{\em reliable} decisions of the symbols in data streams---originating from 
the neighboring sections---to estimate the channel gains 
for the other data streams. Consequently, coupling could not work 
for channel estimation. We use the assignment with the cyclic shift to avoid  
such a defect. 

We refer to the ensemble of all possible STC BICM schemes---induced by the 
randomness of coupled interleavers~(\ref{coupled_interleaver})---as 
the $(M,L,W,Q)$ STC BICM ensemble. 

\subsection{Summary of Transmitter}
The data matrices are generated via $(d_{\mathrm{v}}, d_{\mathrm{c}}, L)$ SC LDPC 
coding and $(M,L,W,Q)$ STC BICM. 
Known binary words $\mathcal{U}_{[-W:0)}=\{\boldsymbol{u}_{l}
\in\mathbb{F}_{2}^{M}:l\in[-W:0)\}$ of length~$M$ are used for the first $W$ 
sections of the coupled interleaver. For the remaining sections, 
$L$ codewords $\mathcal{U}_{[0:L)}=\{\boldsymbol{u}_{l}\in\mathbb{F}_{2}^{M}:
l\in[0:L)\}$ of length~$M$ are generated with the SC LDPC code. 
Then, these $L+W$ binary sequences are interleaved with the coupled  
interleaver. The interleaved bit sequence in each section is mapped to 
$M/Q$ data symbols with unit average power, as presented in 
Section~\ref{sec2B}. In this paper, we only consider $2^{Q}$ quadrature 
amplitude modulation (QAM) with Gray mapping for an even number~$Q$.    
In order to minimize the decoding latency, the codewords are transmitted 
in the same order as that in decoding. 
The overall transmission rate is given by
\begin{equation} \label{rate} 
R = 
\left(
 1 - \frac{T_{\mathrm{tr}}}{T}
\right)
\left(
 1 - \frac{W}{L+W}
\right)QKr,  
\end{equation}
where $r$ denotes the design coding rate, given by (\ref{design_rate}) for  
$(d_{\mathrm{v}}, d_{\mathrm{c}}, L)$ SC LDPC codes. 

\section{Iterative Receivers} \label{section3} 
\subsection{Iterative CE-MUDD} 
\begin{figure}[t]
\begin{center}
\includegraphics[width=\hsize]{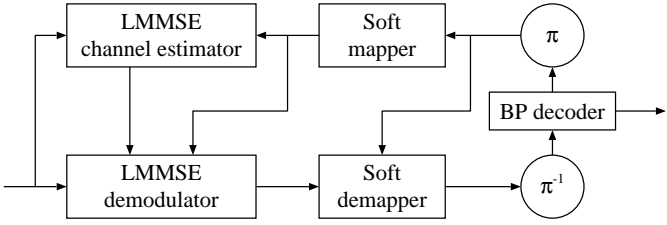}
\end{center}
\caption{
Iterative channel estimation, multiuser detection, and decoding. 
$\pi$ and $\pi^{-1}$ represent the interleaver and the deinterleaver, 
respectively. 
}
\label{fig3}
\end{figure}

The goal of the receiver is to estimate all codewords 
$\mathcal{U}_{[0:L)}$ 
from all received matrices $\mathcal{Y}=\{\boldsymbol{Y}(b):\hbox{all $b$}\}$, 
all pilot matrices $\mathcal{X}_{[1:T_{\mathrm{tr}}]}
=\{\boldsymbol{X}_{[1:T_{\mathrm{tr}}]}(b):\hbox{all $b$}\}$, and 
all known words $\mathcal{U}_{[-W:0)}$. 
We shall derive low-complexity iterative receivers based on approximate BP. 
The iterative receivers consist of the soft mapper, the LMMSE channel 
estimator, the LMMSE demodulator, the soft demapper, and the BP　
decoder, as shown in Fig.~\ref{fig3}. The channel estimator calculates 
approximate a posteriori probability density functions (pdfs) of the channel 
matrices, and sends them to the demodulator. The demodulator 
uses the a posteriori pdfs and the received matrices to calculate approximate a 
posteriori distributions for the data symbols. The a posteriori distributions 
are fed forward to the soft demapper to calculate the log likelihood ratio  
(LLR) for each bit of the codewords. The LLRs are used as the 
a priori LLRs in the BP decoder to calculate extrinsic LLRs. 
The extrinsic LLRs are fed back to the soft mapper and demapper 
in order to refine the initial estimates. After several 
iterations, the BP decoder decodes the codewords on the basis of the a 
posteriori LLRs. In this paper, we refer to iterations in the BP decoder and 
in the CE-MUDD as inner and outer iterations, respectively. 

We hereafter focus on detection of the $k$th data symbol $x_{k,t}(b)$ in symbol 
period~$t$ within a fading block~$b$. For notational convenience, 
we omit the index~$b$. 

\subsection{Soft Mapper}
Suppose that the soft mapper has received $Q$ extrinsic LLRs 
$\mathcal{L}^{\mathrm{dec}}=\{L_{q}^{\mathrm{dec}}:q=1,\ldots,Q\}$ associated with 
the data symbol $x_{k',t'}$ ($k'\neq k$) from the decoder. The a priori 
distribution $P^{\mathrm{dec}}(c_{q})$ of the bit $c_{q}\in\{0,1\}$ corresponding 
to the $q$th LLR $L_{q}^{\mathrm{dec}}$ is given by 
\begin{equation} \label{each_prior} 
P^{\mathrm{dec}}(c_{q}) 
= \frac{e^{(1-2c_{q})L_{q}^{\mathrm{dec}}/2}}
{e^{L_{q}^{\mathrm{dec}}/2}+e^{-L_{q}^{\mathrm{dec}}/2}}. 
\end{equation}
Thus, the a priori distribution of $x_{k',t'}$ to be fed forward is equal to  
\begin{equation} \label{prior}
P(x_{k',t'}=F(\{c_{q}\})) 
= \prod_{q=1}^{Q}P^{\mathrm{dec}}(c_{q}), 
\end{equation}
where $F:\mathbb{F}_{2}^{Q}\rightarrow\mathcal{C}\subset\mathbb{C}$ denotes  
the mapping function. 

Let $\hat{x}_{k',t'}$ and $\sigma_{k',t'}^{2}$ denote the mean and variance 
of $x_{k',t'}$ with respect to (\ref{prior}), respectively. 
As presented in the following sections, the a priori distribution~(\ref{prior}) 
is approximated by a proper complex Gaussian distribution with 
mean~$\hat{x}_{k',t'}$ and variance~$\sigma_{k',t'}^{2}$ to reduce the complexity 
of the channel estimator and the demodulator. Thus, the mean and variance are 
fed forward to the channel estimator and the demodulator, instead of the 
distribution~(\ref{prior}).  

\begin{example}[QPSK] \label{example1}
For QPSK ($Q=2$), suppose that the bits $c_{1}$ and $c_{2}$ are mapped to the 
real and imaginary parts of the data symbol $x_{k',t'}$, respectively. 
The soft decision $\hat{x}_{k',t'}=\hat{x}(\mathcal{L}^{\mathrm{dec}})$ is 
defined as the mean of $x_{k',t'}$ with respect to the a priori 
distribution~(\ref{prior}). The soft decision function $\hat{x}(\cdot)$ is 
explicitly given by 
\begin{equation} 
\hat{x}(\mathcal{L}^{\mathrm{dec}})
= \frac{1}{\sqrt{2}}\tanh\left(
  \frac{L_{1}^{\mathrm{dec}}}{2}
 \right)
+ \frac{\mathrm{i}}{\sqrt{2}}\tanh\left(
  \frac{L_{2}^{\mathrm{dec}}}{2}
 \right). 
\end{equation}

The variance $\sigma_{k',t'}^{2}$ of $x_{k',t'}$ with respect to the 
a priori distribution~(\ref{prior}) is equal to 
$\sigma_{k',t'}^{2}=\sigma^{2}(\mathcal{L}^{\mathrm{dec}})$, with 
\begin{equation}
\sigma^{2}(\mathcal{L}^{\mathrm{dec}}) 
= 1 - |\hat{x}(\mathcal{L}^{\mathrm{dec}})|^{2}. 
\end{equation} 
\end{example}

\begin{example}[16 QAM] \label{example2} 
For 16 QAM ($Q=4$), suppose that the first two bits $c_{1}$ and $c_{2}$ are 
mapped to $\Re[x_{k',t'}]$ with mapping 
$\{(1,1),(0,1),(0,0),(1,0)\}\rightarrow \{-3,-1,1,3\}/\sqrt{10}$, and that 
the last two bits $c_{3}$ and $c_{4}$ are mapped to 
$\Im[x_{k',t'}]$ with the same mapping. The soft decision 
is given by $\hat{x}_{k',t'}=\hat{x}(\mathcal{L}^{\mathrm{dec}})$, 
\begin{IEEEeqnarray}{rl}
\hat{x}(\mathcal{L}^{\mathrm{dec}})
=
& 
\frac{1}{\sqrt{10}}\tanh\left(
 \frac{L_{2}^{\mathrm{dec}}}{2}
\right)\left\{
 2 - \tanh\left(
  \frac{L_{1}^{\mathrm{dec}}}{2}
 \right)
\right\} 
\nonumber \\ 
+
& 
\frac{\mathrm{i}}{\sqrt{10}}\tanh\left(
 \frac{L_{4}^{\mathrm{dec}}}{2}
\right)\left\{
 2 - \tanh\left(
  \frac{L_{3}^{\mathrm{dec}}}{2}
 \right)
\right\}.
\end{IEEEeqnarray}

The variance $\sigma_{k',t'}^{2}$ of $x_{k',t'}$ with respect to the 
a priori distribution~(\ref{prior}) is given by 
$\sigma_{k',t'}^{2}=\sigma^{2}(\mathcal{L}^{\mathrm{dec}})$, with 
\begin{IEEEeqnarray}{rl}
\sigma^{2}(\mathcal{L}^{\mathrm{dec}})
=
& 
1 - \frac{2}{5}\left\{
 \tanh\left(
  \frac{L_{1}^{\mathrm{dec}}}{2}
 \right) 
 + \tanh\left(
  \frac{L_{3}^{\mathrm{dec}}}{2}
 \right) 
\right\} 
\nonumber \\ 
&
- |\hat{x}(\mathcal{L}^{\mathrm{dec}})|^{2}. 
\end{IEEEeqnarray}
\end{example}

\subsection{LMMSE Channel Estimator}
We follow \cite{Takeuchi134} to derive the LMMSE channel estimator. 
Recall that we are focusing on symbol period~$t\in(T_{\mathrm{tr}}:T]$ in a 
fading block. Suppose that the soft decisions 
$\{(\hat{\boldsymbol{X}})_{k',t'}=\hat{x}_{k',t'}:\hbox{$t'\neq t$, all $k'$}\}$ 
and the a priori variances $\{\sigma_{k',t'}^{2}:\hbox{$t'\neq t$, all $k'$}\}$ 
have been provided by the soft mapper. For the training phase 
$t'\in[1:T_{\mathrm{tr}}]$, $\hat{x}_{k',t'}$ and $\sigma_{k',t'}^{2}$ are set to the 
true symbol $x_{k',t'}$ and zero, respectively. 
Let $\boldsymbol{Y}_{\backslash t}$,  
$\boldsymbol{X}_{\backslash t}$, and $\hat{\boldsymbol{X}}_{\backslash t}$ denote 
the matrices obtained by eliminating the $t$th column from the received matrix 
$\boldsymbol{Y}$, the transmitted matrix $\boldsymbol{X}$, and the 
soft decision matrix $\hat{\boldsymbol{X}}$, respectively. 
The channel estimator calculates an approximate a 
posteriori pdf of the channel matrix $\boldsymbol{H}$ given 
$\boldsymbol{Y}_{\backslash t}$, $\hat{\boldsymbol{X}}_{\backslash t}$, and 
$\{\sigma_{k',t'}^{2}:t'\neq t, \hbox{all $k'$}\}$ to be fed 
forward to the demodulator for the $t$th symbol period. Excluding the $t$th 
symbol period stabilizes iterative CE-MUDD, and enables the DE analysis. 

In order to obtain a tractable approximation of the a posteriori pdf, we 
decompose the first term on the right-hand side (RHS) of (\ref{matrix_MIMO}) 
into two terms, 
\begin{equation} \label{MIMO_t} 
\boldsymbol{Y}_{\backslash t} 
= \boldsymbol{H}\hat{\boldsymbol{X}}_{\backslash t} 
+ \boldsymbol{H}(\boldsymbol{X}_{\backslash t} - \hat{\boldsymbol{X}}_{\backslash t})
+ \tilde{\boldsymbol{W}}_{\backslash t}, 
\end{equation}
where $\tilde{\boldsymbol{W}}_{\backslash t}$ is obtained by eliminating the 
$t$th column from the AWGN matrix $\boldsymbol{W}$ and subsequently by 
replacing the noise variance $N_{0}$ with a postulated value $\tilde{N}_{0}$. 
Treating the second term exactly results in nonlinear channel estimator, 
so that the term given $\hat{\boldsymbol{X}}_{\backslash t}$ is approximated 
by the CSCG random matrix with the same covariance as the original one. 
The second term 
$\boldsymbol{H}(\boldsymbol{X}_{\backslash t} - \hat{\boldsymbol{X}}_{\backslash t})$ 
has independent rows with the covariance matrix 
\begin{equation} \label{Sigma_x}
\boldsymbol{\Sigma}_{\backslash t}^{\mathrm{ch}}
=\mathrm{diag}\{\sigma_{t'}^{2}:t'\neq t\}, 
\quad 
\sigma_{t'}^{2} 
= \frac{1}{K}\sum_{k'=1}^{K}\sigma_{k',t'}^{2}. 
\end{equation}
Calculating the a posteriori pdf of $\boldsymbol{H}$ under this approximation, 
we obtain 
\begin{IEEEeqnarray}{rl}
&
p(\boldsymbol{H}|\boldsymbol{Y}_{\backslash t},
\hat{\boldsymbol{X}}_{\backslash t})
\nonumber \\ 
=
& 
\frac{1}{(\pi^{K}\det\boldsymbol{\Xi}_{\backslash t}^{\mathrm{ch}})^{N}}
e^{
-\mathrm{Tr}\left\{
 (\boldsymbol{\Xi}_{\backslash t}^{\mathrm{ch}})^{-1}
 (\boldsymbol{H}-\hat{\boldsymbol{H}}_{\backslash t})^{\mathrm{H}}
 (\boldsymbol{H}-\hat{\boldsymbol{H}}_{\backslash t})
\right\}
}, \label{posterior_H} 
\end{IEEEeqnarray}
where the LMMSE estimate $\hat{\boldsymbol{H}}_{\backslash t}\in
\mathbb{C}^{N\times K}$ and the a posteriori covariance matrix 
$\boldsymbol{\Xi}_{\backslash t}^{\mathrm{ch}}$ are given by  
\begin{equation} \label{LMMSE_H} 
\hat{\boldsymbol{H}}_{\backslash t} 
= \boldsymbol{Y}_{\backslash t}(\boldsymbol{\Sigma}_{\backslash t}^{\mathrm{ch}}
 +\tilde{N}_{0}\boldsymbol{I}_{T-1})^{-1}\hat{\boldsymbol{X}}_{\backslash t}^{\mathrm{H}}
 \boldsymbol{\Xi}_{\backslash t}^{\mathrm{ch}}, 
\end{equation} 
\begin{equation} \label{posterior_covariance} 
\boldsymbol{\Xi}_{\backslash t}^{\mathrm{ch}} 
= \left\{
 K\boldsymbol{I}_{K} 
 + \hat{\boldsymbol{X}}_{\backslash t}(\boldsymbol{\Sigma}_{\backslash t}^{\mathrm{ch}}
 +\tilde{N}_{0}\boldsymbol{I}_{T-1})^{-1}\hat{\boldsymbol{X}}_{\backslash t}^{\mathrm{H}}
\right\}^{-1}, 
\end{equation}
respectively. In (\ref{posterior_H}), we have omitted the conditioning with 
respect to $\boldsymbol{\Sigma}_{\backslash t}^{\mathrm{ch}}$ for notational 
convenience. 
See Appendix~\ref{appendix_A} for an efficient calculation of 
(\ref{posterior_covariance}) for all $t$. 

Expression~(\ref{posterior_H}) implies that the approximate a 
posteriori pdf of $\boldsymbol{H}$ is a proper complex Gaussian pdf. 
Sending (\ref{posterior_H}) is equivalent to feeding the LMMSE 
estimate~(\ref{LMMSE_H}) and the a posteriori  
covariance~(\ref{posterior_covariance}) forward to the demodulator. 

\begin{remark} \label{remark1}
Using a mismatched value $\tilde{N}_{0}\neq N_{0}$ should result in no 
improvement in the decoding threshold, since the LMMSE channel estimator 
with $\tilde{N}_{0}=N_{0}$ can achieve the nonlinear minimum mean-squared 
error (MMSE) performance in the large-system limit~\cite{Takeuchi134}. 
Thus, we assume $\tilde{N}_{0}=N_{0}$ in the large-system analysis after taking 
the limit $M\to\infty$. In order to avoid error 
propagation for finite-sized systems, on the other hand, we may use a 
mismatched value $\tilde{N}_{0}\neq N_{0}$ in numerical simulations. 
By definition, the covariance matrix~(\ref{Sigma_x}) vanishes as the a priori 
variances $\{\sigma_{k,t}^{2}\}$ of the data symbols tend to zero. This 
typically occurs in the high SNR regime, in which the inverse matrix 
$(\boldsymbol{\Sigma}_{\backslash t}^{\mathrm{ch}}+
\tilde{N}_{0}\boldsymbol{I}_{T-1})^{-1}$ in (\ref{LMMSE_H}) and 
(\ref{posterior_covariance}) diverges when $\tilde{N}_{0}=N_{0}$ is assumed. 
This divergence may result in error propagation for finite-sized systems. 
In order to circumvent the divergence, we may use a fixed value 
$\tilde{N}_{0}>0$ in the high SNR regime, 
whereas we set $\tilde{N}_{0}=N_{0}$ in the other SNR regime. 
Note that a mismatched value $\tilde{N}_{0}$ is used only for the channel 
estimator. We always assume $\tilde{N}_{0}=N_{0}$ in the demodulator, since 
such a divergence does not occur for the demodulator as long as the channel 
estimator uses a mismatched value $\tilde{N}_{0}\neq N_{0}$.  
\end{remark}

\subsection{LMMSE Demodulator} 
We are focusing on the $k$th data symbol $x_{k,t}$ in symbol period~$t$. 
Suppose that the soft decisions $\{\hat{x}_{k',t}\}$ 
and the a priori variances $\{\sigma_{k',t}^{2}\}$ have been provided by the 
soft mapper, and that the a posteriori pdf 
$p(\boldsymbol{H}|\boldsymbol{Y}_{\backslash t},\hat{\boldsymbol{X}}_{\backslash t})$ 
sent by the channel estimator is used as the a priori pdf of the channel 
matrix. 

In order to obtain a tractable a posteriori distribution of the data symbol 
$x_{k,t}$, we shall make the same approximation as in the derivation of the 
LMMSE channel estimator.  We first decompose the first term on the RHS of 
(\ref{MIMO}) into two terms, 
\begin{equation}
\boldsymbol{y}_{t} 
= \hat{\boldsymbol{H}}_{\backslash t}\boldsymbol{x}_{t} 
+ (\boldsymbol{H} - \hat{\boldsymbol{H}}_{\backslash t})\boldsymbol{x}_{t} 
+ \boldsymbol{w}_{t}, 
\end{equation}
where the LMMSE channel estimate $\hat{\boldsymbol{H}}_{\backslash t}$ is 
given by (\ref{LMMSE_H}). We next approximate the second term by a CSCG 
term with covariance $\zeta_{t}\boldsymbol{I}_{N}$, given by 
\begin{equation} \label{eta} 
\zeta_{t} = \mathrm{Tr}\left(
 \boldsymbol{\Xi}_{\backslash t}^{\mathrm{ch}}
 \mathrm{diag}\{\sigma_{k',t}^{2}+|\hat{x}_{k',t}|^{2}:\hbox{for all $k'$}\}
\right), 
\end{equation}  
where $\boldsymbol{\Xi}_{\backslash t}^{\mathrm{ch}}$ is given by 
(\ref{posterior_covariance}). 
The obtained approximate MIMO channel can be regarded as 
a MIMO channel with the {\em known} channel matrix 
$\hat{\boldsymbol{H}}_{\backslash t}=(\hat{\boldsymbol{h}}_{1,\backslash t},\ldots,
\hat{\boldsymbol{h}}_{K,\backslash t})$ and an AWGN vector with covariance 
$(\zeta_{t}+N_{0})\boldsymbol{I}_{N}$. In order to obtain the LMMSE 
estimator of $x_{k,t}$, we approximate the a priori distributions of the 
other data symbols $x_{k',t}$ by proper complex Gaussian distributions with 
mean $\hat{x}_{k',t}$ and covariance $\sigma_{k',t}^{2}$ for all $k'\neq k$, 
whereas the a priori distribution of $x_{k,t}$ is approximated by the CSCG 
distribution with unit variance. 
As shown in \cite{Takeuchi132}, the approximate a posteriori pdf of $x_{k,t}$ 
is evaluated as  
\begin{equation} \label{posterior_LMMSE} 
p(x_{k,t}|\boldsymbol{y}_{t},\hat{\boldsymbol{H}}_{\backslash t},\{\hat{x}_{k',t}\}) 
= \frac{1}{\pi\xi_{k,t}^{\mathrm{dem}}}\exp\left(
 - \frac{|x_{k,t} - \bar{x}_{k,t}|^{2}}{\xi_{k,t}^{\mathrm{dem}}} 
\right), 
\end{equation}
where the LMMSE estimate $\bar{x}_{k,t}$ and its mean-squared error (MSE)  
$\xi_{k,t}^{\mathrm{dem}}$ are given by  
\begin{equation} \label{LMMSE_x} 
\bar{x}_{k,t} 
= \xi_{k,t}^{\mathrm{dem}}\hat{\boldsymbol{h}}_{k,\backslash t}^{\mathrm{H}}
 \boldsymbol{\Xi}_{\backslash k,t}^{\mathrm{dem}}\left(
 \boldsymbol{y}_{t} - \sum_{k'\neq k}\hat{\boldsymbol{h}}_{k',\backslash t}
 \hat{x}_{k',t}
\right), 
\end{equation}
\begin{equation} \label{xi}
\xi_{k,t}^{\mathrm{dem}} 
= \left(
 1 + \hat{\boldsymbol{h}}_{k,\backslash t}^{\mathrm{H}}
 \boldsymbol{\Xi}_{\backslash k,t}^{\mathrm{dem}}
 \hat{\boldsymbol{h}}_{k,\backslash t}
\right)^{-1}, 
\end{equation}
respectively, In these expressions, 
$\boldsymbol{\Xi}_{\backslash k,t}^{\mathrm{dem}}$ is defined as  
\begin{equation} \label{MSE} 
\boldsymbol{\Xi}_{\backslash k,t}^{\mathrm{dem}}  
= \left(
 (N_{0}+\zeta_{t})\boldsymbol{I}_{N} 
 + \sum_{k'\neq k}\sigma_{k',t}^{2}\hat{\boldsymbol{h}}_{k',\backslash t}
 \hat{\boldsymbol{h}}_{k',\backslash t}^{\mathrm{H}}
\right)^{-1}. 
\end{equation}
In (\ref{posterior_LMMSE}), we have omitted the conditioning with respect to 
$\zeta_{t}$ and $\{\sigma_{k',t}^{2}:k'\neq k\}$. 
Since (\ref{eta}) is independent of the index~$k$, it is possible to 
calculate (\ref{MSE}) for all $k$ efficiently, by using the same method as 
in Appendix~\ref{appendix_A}. 

\begin{remark}
In order to reduce the complexity of the demodulator, we have used the 
feedback information $\hat{x}_{k,t}$ and $\sigma_{k,t}^{2}$ 
about the $k$th data symbol $x_{k,t}$ in the definition of (\ref{eta}). 
This violates the update rule of BP in which the feedback information about 
$x_{k,t}$ should not be used in its detection. However, this influence will 
be shown to vanish in the large-system limit. As a result, the proposed LMMSE 
demodulator in that limit reduces to the true LMMSE demodulator in which 
$\hat{x}_{k,t}$ and $\sigma_{k,t}^{2}$ in (\ref{eta}) are replaced by $0$ and 
$1$, respectively. 
\end{remark}

\subsection{Soft Demapper} 
The soft demapper sends the LLR $L_{q}^{\mathrm{dem}}$ of the $q$th bit $c_{q}$ 
for the data symbol $x_{k,t}$ to the decoder for $q=1,\ldots,Q$. 
The extrinsic probability $P^{\mathrm{dem}}(c_{q})$ of $c_{q}$ is given by 
\begin{IEEEeqnarray}{rl}
P^{\mathrm{dem}}(c_{q}) 
\propto \sum_{\{c_{q'}\}\backslash c_{q}}
&
p(x_{k,t}=F(\{c_{q'}\})|\boldsymbol{y}_{t},\hat{\boldsymbol{H}}_{\backslash t},
\{\hat{x}_{k',t}\}) 
\nonumber \\ 
&\cdot
\prod_{q'\neq q}P^{\mathrm{dec}}(c_{q'}). \label{ext_probability} 
\end{IEEEeqnarray} 
In (\ref{ext_probability}), $F$ denotes the mapping function, and the pdf 
$p(x_{k,t}|\boldsymbol{y}_{t},\hat{\boldsymbol{H}}_{\backslash t},\{\hat{x}_{k',t}\})$ 
sent from the demodulator is given by (\ref{posterior_LMMSE}). Furthermore,  
$P_{\mathrm{dec}}(c_{q'})$ represents the a priori probability~(\ref{each_prior}). 
The summation in (\ref{ext_probability}) is 
taken over all possible binary sequences $(c_{1},\ldots,c_{Q})$ with $c_{q}$ 
fixed. Then, the extrinsic LLR $L_{q}^{\mathrm{dem}}$ to be 
fed forward is defined as 
\begin{equation}
L_{q}^{\mathrm{dem}} \label{ext_LLR} 
= \ln\frac{P^{\mathrm{dem}}(c_{q}=0)}{P^{\mathrm{dem}}(c_{q}=1)}.  
\end{equation}

\subsection{Sliding-Window Schedule} 
The performance of the iterative CE-MUDD depends on message-passing schedules.  
We use a sliding-window (SW) schedule with a window size of $W_{\mathrm{SW}}$ 
sections. The decoding window runs in the ascending order of 
sections~$l=0,\ldots,L-1$, because of the termination 
structure~(\ref{base_matrix}) of the SC LDPC code. 
In stage~$l\in[0:L-W_{\mathrm{SW}}]$ of the SW decoding, $W_{\mathrm{SW}}$ codewords 
in sections~$[l:l+W_{\mathrm{SW}})$ are decoded on the basis of the decoding 
results in the preceding stages and of the received signals in the 
$(W_{\mathrm{SW}}+2W)$ sections $[l-W:l+W_{\mathrm{SW}}+W)$, associated with 
the $W_{\mathrm{SW}}$ codewords, while the received signals in the sections 
$[l-W:L)$ are used for the last stages $l\in[L-W_{\mathrm{SW}}-W:L-W_{\mathrm{SW}}]$. 
Thus, the decoding latency is $O((W_{\mathrm{SW}}+2W)M)$, and independent of the 
number $L$ of sections. 

In each stage, we regard the four parts separated by the interleaver and 
the deinterleaver---shown in Fig.~\ref{fig3}---as one effective demodulator 
{\em without} inner iterations. Thus, messages are updated between the BP 
decoder with inner iterations and the effective demodulator. 
We use the parallel schedule in the outer iterations between the effective 
demodulator and the decoder. In the parallel schedule, all data symbols 
in $(W_{\mathrm{SW}}+2W)$ sections are demodulated and then fed forward to 
the decoder. The decoder use the received messages to perform iterative 
decoding of the $W_{\mathrm{SW}}$ codewords on the basis of on-demand check node 
updating schedule~\cite{Lentmaier102} with inner iterations~$I$, which 
is the total number of updating the message of each variable node in 
one decoding. Thus, the total number of updating each variable node is $IJ$ 
in each stage of the SW schedule, in which $J$ is the number of outer 
iterations between the effective demodulator and the decoder in each stage.  
Furthermore, the corresponding total number is equal to $W_{\mathrm{SW}}IJ$ 
in one CE-MUDD for the bulk region of sections.  

In one round of stage~$l$ for the SW decoding with the on-demand check node 
updating schedule, the variable nodes in sections~$l'$ are updated in the 
ascending order $l'=l,\ldots,l+W_{\mathrm{SW}}-1$. All variable nodes in 
each section are updated simultaneously, after updating all check nodes that 
are directly connected to the variable nodes. Such $I$ rounds are repeated 
in each stage of the BP decoder. See \cite{Lentmaier102} for the details.

\section{Density Evolution Analysis} \label{section4}
\subsection{Large System Analysis} 
The DE analysis is presented in the large-system limit after taking the 
long code-length limit $M\to\infty$. 
In the large-system limit, $N$, $K$, $T$, and $T_{\mathrm{tr}}$ tend to infinity 
at the same rate. We note that each codeword is transmitted over many fading 
blocks since $M$ tends to infinity with coherence time $T$ fixed. 
In this sense, the limits in this paper correspond to fast fading.  

As noted in Remark~\ref{remark1}, we assume that the noise value 
$\tilde{N}_{0}$ postulated in the channel estimator is equal to the true one 
$N_{0}$ in the DE analysis. 
We evaluate the average performance over all possible information bits, the 
$(d_{\mathrm{v}},d_{\mathrm{c}},L)$ ensemble of SC LDPC codes, and the 
$(M,L,W,Q)$ ensemble of STC BICM. For given information bits, an SC LDPC 
code, and STC BICM, each soft decision $\hat{x}_{k,t}$ should be biased 
toward the true data symbol $x_{k,t}$ as the iteration proceeds. However, the 
average $\mathbb{E}[\hat{x}_{k,t}]$ over all possible randomness is equal to 
zero since the data symbol $x_{k,t}$ should appear 
on the constellation points $\mathcal{C}$ uniformly. Furthermore, the 
assumption of random bit-interleaving implies that the soft decisions and 
the a priori variances $\{\hat{x}_{k,t},\sigma_{k,t}^{2}\}$ are independent 
for all $k$ and $t$ in the limit $M\to\infty$. We use these asymptotic 
properties to analyze the channel estimator and the demodulator.  

\subsection{LMMSE Channel Estimator}
We first present the large-system analysis for the LMMSE channel 
estimator. Let us focus on symbol period~$t$ within a fading block 
in section~$l$. Let $\xi_{\backslash t}(l)$ denote the average 
MSE of the LMMSE channel estimation in section~$l$ over the codewords and 
the ensemble of STC BICM, 
\begin{equation} \label{average_MSE} 
\xi_{\backslash t}(l) 
= \mathbb{E}\left[
 \mathrm{Tr}(\boldsymbol{\Xi}_{\backslash t}^{\mathrm{ch}})
\right], 
\end{equation}
where $\boldsymbol{\Xi}_{\backslash t}^{\mathrm{ch}}$ is given by 
(\ref{posterior_covariance}). 
We note that (\ref{average_MSE}) is $O(1)$ in the large-system limit. 
The average MSE is characterized by 
the average squared soft decision $\mathbb{E}[|\hat{x}_{k',t'}|^{2}]$. 
From the construction of the STC BICM, $\mathbb{E}[|\hat{x}_{k',t'}|^{2}]$ in 
section~$l$ is given by 
\begin{equation} \label{squared_decision} 
\hat{x}_{\mathrm{dec}}^{2}(l) 
=\frac{1}{2W+1}\sum_{w=-W}^{W}\mathbb{E}\left[
 |\hat{x}(\mathcal{L}_{f_{l}(w)}^{\mathrm{dec}})|^{2}
\right], 
\end{equation}
with 
\begin{equation} \label{index_function}
f_{l}(w) = 
\left\{
\begin{array}{cl}
l+w & \hbox{for $l+w\in[-W:L)$} \\
l & \hbox{otherwise.}
\end{array} 
\right.
\end{equation} 
In (\ref{squared_decision}), $\mathcal{L}_{l}^{\mathrm{dec}}$ denotes the set of 
$Q$ independent and identically distributed (i.i.d.) random variables that 
represent the LLRs fed back from the decoder in section~$l$. Furthermore, 
the soft decision $\hat{x}(\mathcal{L}^{\mathrm{dec}})$ is the mean of the 
data symbol with respect to the a priori distribution~(\ref{prior}) defined 
via the LLRs $\mathcal{L}^{\mathrm{dec}}$. See Examples \ref{example1} and 
\ref{example2} for QPSK and $16$ QAM, respectively. The expectation 
in (\ref{squared_decision}) is taken over the distribution of the 
LLRs $\mathcal{L}_{f_{l}(w)}^{\mathrm{dec}}$, which will be analyzed shortly.   

It is shown that the average MSE~(\ref{average_MSE}) is given via the 
following function in the large-system limit:  
\begin{equation} \label{asymptotic_MSE} 
\xi(\sigma_{\mathrm{tr}}^{2},\sigma_{\mathrm{c}}^{2};l) 
= \left(
 1 + \frac{T_{\mathrm{tr}}}{K\sigma_{\mathrm{tr}}^{2}}
 + \frac{(T-T_{\mathrm{tr}}-1)\hat{x}_{\mathrm{dec}}^{2}(l)}{K\sigma_{\mathrm{c}}^{2}}
\right)^{-1}. 
\end{equation}
See \cite{Takeuchi134} for the operational meaning of (\ref{asymptotic_MSE}). 

\begin{proposition} \label{proposition1} 
Focus on a fading block in section~$l$. 
Suppose that $\hat{x}_{\mathrm{dec}}^{2}(l)$ is given by (\ref{squared_decision}) 
via the feedback information from the decoder in an iteration round, and that 
$(\sigma_{\mathrm{tr}}^{2},\sigma_{\mathrm{c}}^{2})$ is the solution to the 
coupled fixed-point (FP) equations 
\begin{equation} \label{FP_tr}
\sigma_{\mathrm{tr}}^{2} 
= N_{0} + \xi(\sigma_{\mathrm{tr}}^{2},\sigma_{\mathrm{c}}^{2};l), 
\end{equation}
\begin{equation} \label{FP_c} 
\sigma_{\mathrm{c}}^{2} 
= N_{0} + 1 - \hat{x}_{\mathrm{dec}}^{2}(l) 
+ \hat{x}_{\mathrm{dec}}^{2}(l)\xi(\sigma_{\mathrm{tr}}^{2},\sigma_{\mathrm{c}}^{2};l), 
\end{equation}
where $\hat{x}_{\mathrm{dec}}^{2}(l)$ and 
$\xi(\sigma_{\mathrm{tr}}^{2},\sigma_{\mathrm{c}}^{2};l)$ are given by 
(\ref{squared_decision}) and (\ref{asymptotic_MSE}), respectively. Then, 
the difference between the two MSEs~(\ref{average_MSE}) and 
(\ref{asymptotic_MSE}) converges to zero in the large-system limit after 
taking the limit $M\to\infty$.  
\end{proposition}
\begin{IEEEproof}
See Appendix~\ref{proof_proposition1}. 
\end{IEEEproof}

The quantity~(\ref{asymptotic_MSE}) has a well-defined limit as $K$, 
$T$, and $T_{\mathrm{tr}}$ tend to infinity with their ratios $K/T$ and 
$T_{\mathrm{tr}}/T$ kept constant. In this paper, 
(\ref{asymptotic_MSE}) is used as an approximation of the average 
MSE~(\ref{average_MSE}) for finite-sized systems. 
It was numerically demonstrated in \cite{Takeuchi134} that 
(\ref{asymptotic_MSE}) is a good approximation for small MIMO systems.  

\subsection{LMMSE Demodulator} 
We next present the large-system analysis for the LMMSE demodulator. 
Let us define the equivalent channel between the mapper and the demapper as 
\begin{IEEEeqnarray}{r} 
p(\tilde{x}_{k,t}|x_{k,t},\{\hat{x}_{k',t}\}) 
= \mathbb{E}_{\boldsymbol{H},\hat{\boldsymbol{H}}_{\backslash t}}\biggl[
\int p(\boldsymbol{y}_{t}|\boldsymbol{H},x_{k,t})
\nonumber \\ \cdot 
p(x_{k,t}=\tilde{x}_{k,t}|\boldsymbol{y}_{t},
\hat{\boldsymbol{H}}_{\backslash t},\{\hat{x}_{k',t}\})
d\boldsymbol{y}_{t}
\biggr]. \label{equivalent_channel} 
\end{IEEEeqnarray}
In (\ref{equivalent_channel}), the pdf 
$p(\boldsymbol{y}_{t}|\boldsymbol{H},x_{k,t})$ represents the MIMO 
channel~(\ref{MIMO}) in symbol period~$t$ marginalized over the data symbols 
$\{x_{k',t}:k'\neq k\}$ with the exception of $x_{k,t}$. The a posteriori pdf 
$p(x_{k,t}|\boldsymbol{y}_{t},\hat{\boldsymbol{H}}_{\backslash t},\{\hat{x}_{k',t}\})$ 
is given by (\ref{posterior_LMMSE}). The expectation 
$\mathbb{E}_{\boldsymbol{H},\hat{\boldsymbol{H}}_{\backslash t}}[\cdot]$ is taken over the 
joint distribution of the channel matrix $\boldsymbol{H}$ and its LMMSE 
estimate~(\ref{LMMSE_H}) induced from (\ref{MIMO_t}), as well as over 
all possible codewords and interleavers. It is shown that the equivalent 
channel converges to that for an interference-free AWGN channel in the 
large-system limit. 

We focus on section~$l$, and define the AWGN channel as 
\begin{equation} \label{AWGN} 
z_{l} 
= \sqrt{1-\xi(\sigma_{\mathrm{tr}}^{2},\sigma_{\mathrm{c}}^{2};l)}x_{l} 
+ n_{l}, 
\; n_{l}\sim\mathcal{CN}(0,\sigma_{\mathrm{dem}}^{2}(l)),  
\end{equation}
with $x_{l}$ denoting the input symbol for section~$l$. In (\ref{AWGN}), 
$\xi(\sigma_{\mathrm{tr}}^{2},\sigma_{\mathrm{c}}^{2};l)$ defined by 
(\ref{asymptotic_MSE}) is given via the solution to the coupled FP 
equations~(\ref{FP_tr}) and (\ref{FP_c}). 
The equivalent channel for the AWGN channel~(\ref{AWGN}) is defined as 
\begin{equation} \label{equivalent_AWGN} 
p(\tilde{x}_{l}|x_{l}) 
= \int p(x_{l}=\tilde{x}_{l} | z_{l})p(z_{l}| x_{l})dz_{l}.  
\end{equation}
In (\ref{equivalent_AWGN}), $p(z_{l}|x_{l})$ denotes the AWGN 
channel~(\ref{AWGN}). Furthermore, $p(x_{l} | z_{l})$ is 
the a posteriori pdf of the input symbol with the Gaussian a priori pdf 
$x_{l}\sim\mathcal{CN}(0,1)$. 

\begin{proposition} \label{proposition2} 
Focus on section~$l$. Suppose that 
$\xi_{l}\equiv\xi(\sigma_{\mathrm{tr}}^{2},\sigma_{\mathrm{c}}^{2};l)$ defined by 
(\ref{asymptotic_MSE}) is given via the solution to the coupled 
FP equations~(\ref{FP_tr}) and (\ref{FP_c}) in Proposition~\ref{proposition1}, 
and that $\sigma_{\mathrm{dem}}^{2}(l)$ is the solution $\sigma_{\mathrm{dem}}^{2}$ 
to the following FP equation 
\begin{equation} \label{FP} 
\sigma_{\mathrm{dem}}^{2} 
= \frac{K}{N}\left\{
 N_{0} + \xi_{l} + \frac{1-\xi_{l}}{2W+1}
 \sum_{w=-W}^{W}\mathrm{MSE}_{f_{l}(w)}(\sigma_{\mathrm{dem}}^{2})  
\right\}, 
\end{equation}
with 
\begin{equation} \label{LMMSE_MSE} 
\mathrm{MSE}_{l}(\sigma_{\mathrm{dem}}^{2})  
= \mathbb{E}\left[
 \frac{\sigma^{2}(\mathcal{L}_{l}^{\mathrm{dec}})\sigma_{\mathrm{dem}}^{2}}
 {(1-\xi_{l})\sigma^{2}(\mathcal{L}_{l}^{\mathrm{dec}}) + \sigma_{\mathrm{dem}}^{2}}
\right]. 
\end{equation}
In (\ref{FP}) and (\ref{LMMSE_MSE}), $f_{l}(w)$ is given by 
(\ref{index_function}). Furthermore, 
$\sigma^{2}(\mathcal{L}_{l}^{\mathrm{dec}})$ is the variance of the data symbol 
that has the a priori distribution~(\ref{prior}) defined via the LLRs 
$\mathcal{L}_{l}^{\mathrm{dec}}$. 
Then, the equivalent channel~(\ref{equivalent_channel}) converges to 
(\ref{equivalent_AWGN}) for the AWGN channel in the large-system limit after 
taking the long code-length limit, i.e.,  
for almost all realizations of $\{\hat{x}_{k',t}\}$ 
\begin{equation}
p(\tilde{x}_{k,t}|x_{k,t},\{\hat{x}_{k',t}\}) 
- p(\tilde{x}_{l}=\tilde{x}_{k,t}|x_{l}=x_{k,t}) \to 0. 
\end{equation}
\end{proposition} 
\begin{IEEEproof}
Repeat the derivation of Proposition~\ref{proposition1}. 
\end{IEEEproof}

The function $\sigma^{2}(\mathcal{L}^{\mathrm{dec}})$ is given in 
Examples \ref{example1} and \ref{example2} for QPSK and $16$ QAM, 
respectively. 
Proposition~\ref{proposition2} implies that the equivalent channel between 
the mapper and the demapper reduces to the AWGN channel in the large-system 
limit. Thus, we can apply the DE analysis of conventional 
BICM~\cite{Fabregas08} to evaluating the expectation in (\ref{LMMSE_MSE}) 
over the LLRs $\{\mathcal{L}_{l}^{\mathrm{dec}}\}$. 

\subsection{Soft Demapper}
We focus on the soft demapper in section$~l$, and analyze the LLR distribution 
that is fed forward to the decoders in sections~$l'=f_{l}(w)$, given by 
(\ref{index_function}), for $w=-W,\ldots,W$. 
In Proposition~\ref{proposition2} we have shown that the equivalent channel 
for section~$l$ reduces to the complex AWGN channel~(\ref{AWGN}). 
Let $P_{l\rightarrow l'}^{\mathrm{dem}}(c_{q})$ denote the extrinsic probability of 
the $q$th bit $c_{q}$ for the data symbol $x_{l}$,  
\begin{equation} \label{ext_probability_AWGN} 
P_{l\rightarrow l'}^{\mathrm{dem}}(c_{q}) 
\propto \sum_{\{c_{q'}\}\backslash c_{q}}p(z_{l}|x_{l}=F(\{c_{q'}\}))
\prod_{q'\neq q}P_{l'}^{\mathrm{dec}}(c_{q'}). 
\end{equation}
In (\ref{ext_probability_AWGN}), $P_{l'}^{\mathrm{dec}}(c_{q'})$ denotes the a 
priori probability fed back from the $l'$th decoder, which is defined in 
the same manner as in (\ref{each_prior}). 
Furthermore, the pdf $p(z_{l}|x_{l})$ represents the 
AWGN channel~(\ref{AWGN}) with $F$ denoting the mapping function. 
The extrinsic probability~(\ref{ext_probability_AWGN}) sent to the $l'$th 
decoder depends only on the feedback information from the same decoder, 
since each data symbol originates from the same codeword. 

From Proposition~\ref{proposition2}, the LLRs~(\ref{ext_LLR}) sent from 
the demapper in section~$l$ to the $l'$th decoder are statistically 
equivalent to the LLRs 
\begin{equation} \label{ext_LLR_AWGN}
L_{l\rightarrow l',q}^{\mathrm{dem}} 
= \ln\frac{P_{l\rightarrow l'}^{\mathrm{dem}}(c_{q}=0)}
{P_{l\rightarrow l'}^{\mathrm{dem}}(c_{q}=1)},  
\quad q=1,\ldots,Q, 
\end{equation}
in the large-system limit. The RHS is a random variable that depends on 
the received signal $z_{l}$ and the LLRs fed back from the $l'$th decoder. 

The distributions of the LLRs~(\ref{ext_LLR_AWGN}) are Gaussian-distributed 
for QPSK ($Q=2$). Furthermore, it is possible to estimate them via numerical 
sampling, although the exact distributions for $Q>2$ have intractable 
expressions. Since the use of Gray mapping has been 
assumed, the LLRs~(\ref{ext_LLR_AWGN}) are classified into statistically 
equivalent two groups for the in-phase and quadrature components. 
The distribution sent to the decoder is the mixture of the distributions of 
the $Q/2$ LLRs in each group. In order to simplify the analysis of the 
demapper, we approximate the mixture distribution by a Gaussian distribution. 
Under the Gaussian approximation, it is sufficient to estimate the average 
entropy $h_{l\rightarrow l'}^{\mathrm{dem}\rightarrow\mathrm{dec}}$ of 
(\ref{ext_probability_AWGN}) over all $q$ via numerical sampling. 

The Gaussian approximation might be too simple to characterize the properties 
of mixture distributions for $Q>2$, although it is a popular approximation 
in the literature. It is beyond the scope of this 
paper to construct more sophisticated approximations. 

Under the Gaussian approximation~\cite{Brink01}, the LLR distribution is 
approximated by the real Gaussian distribution with 
mean~$m_{l\rightarrow l'}^{\mathrm{dem}\rightarrow\mathrm{dec}}$ and 
variance~$2m_{l\rightarrow l'}^{\mathrm{dem}\rightarrow\mathrm{dec}}$ when the 
corresponding true bit is zero. Otherwise, it is approximated by 
$\mathcal{N}(-m_{l\rightarrow l'}^{\mathrm{dem}\rightarrow\mathrm{dec}}, 
2m_{l\rightarrow l'}^{\mathrm{dem}\rightarrow\mathrm{dec}})$. 
We define the function $h_{l\rightarrow l'}^{\mathrm{dem}\rightarrow\mathrm{dec}}
=\psi(m_{l\rightarrow l'}^{\mathrm{dem}\rightarrow\mathrm{dec}})$ to specify the 
relationship between the mean $m_{l\rightarrow l'}^{\mathrm{dem}\rightarrow\mathrm{dec}}$ 
and the entropy~$h_{l\rightarrow l'}^{\mathrm{dem}\rightarrow\mathrm{dec}}$. 
\begin{equation} \label{psi} 
\psi(m) = \int_{\mathbb{R}}S\left(
 \frac{\mathrm{e}^{L/2}}{\mathrm{e}^{L/2}+\mathrm{e}^{-L/2}}
\right)
\frac{1}{\sqrt{4\pi m}}\mathrm{e}^{-\frac{(L-m)^{2}}{4m}}dL, 
\end{equation}
where $S(p)$ denotes the binary entropy function 
\begin{equation}
S(p) = -p\log_{2}p - (1-p)\log_{2}(1-p). 
\end{equation}
The function~$\psi(m)$ is regarded as the 
average entropy of a binary random variable characterized by a 
Gaussian-distributed LLR~$L$ with mean~$m$ and variance $2m$. 
Since (\ref{psi}) is monotonically decreasing, the inverse function 
$\psi^{-1}$ exists.  

\subsection{BP Decoder} \label{sec4-E}
We shall analyze the BP decoder for the $(d_{\mathrm{v}},d_{\mathrm{c}},L)$ ensemble 
of SC LDPC codes under the Gaussian approximation. 
We have shown that the analysis of the $l$th decoder reduces to that of 
the decoder with the information 
$\{h_{l'\rightarrow l}^{\mathrm{dem}\rightarrow\mathrm{dec}}:l'=f_{l}(w), w\in[-W:W]\}$ 
from the demapper. 

Without loss of generality, transmission of all-zero codewords is assumed. 
We focus on section~$l$, and analyze the variable-to-check message update. 
Suppose that the variable nodes in section~$l$ have received the entropy 
$h_{l+w\rightarrow l}^{\mathrm{c}\rightarrow\mathrm{v}}$ from the check nodes in 
section~$l+w$ for $w\in[0:d_{\mathrm{v}})$ in an inner iteration. 
We approximate the pdf of the LLRs emitted from the check nodes 
by the Gaussian pdf with mean 
$\psi^{-1}(h_{l+w\rightarrow l}^{\mathrm{c}\rightarrow\mathrm{v}})$ and 
variance $2\psi^{-1}(h_{l+w\rightarrow l}^{\mathrm{c}\rightarrow\mathrm{v}})$. 
From the construction of the $(d_{\mathrm{v}}, d_{\mathrm{c}}, L)$ ensemble of 
protograph-based SC LDPC codes, each variable node in section $l$ emits  
the sum of LLRs sent from  a demapper and $d_{\mathrm{v}}-1$ check nodes. 
Thus, the average entropy $h_{l\rightarrow l+w}^{\mathrm{v}\rightarrow\mathrm{c}}$ passed 
from the variable nodes in section~$l$ to the check nodes in section~$l+w$ 
is given by 
\begin{IEEEeqnarray}{rl}  
h_{l\rightarrow l+w}^{\mathrm{v}\rightarrow\mathrm{c}} =& 
\frac{1}{2W+1}\sum_{w'=-W}^{W}\psi\Biggl(
\psi^{-1}(h_{f_{l}(w')\rightarrow l}^{\mathrm{dem}\rightarrow\mathrm{dec}}) 
\nonumber \\ 
&+\left. 
 \sum_{w''\in[0:d_{\mathrm{v}}), w''\neq w}
 \psi^{-1}(h_{l+w''\rightarrow l}^{\mathrm{c}\rightarrow\mathrm{v}})
\right). \label{entropy_v}
\end{IEEEeqnarray}
In (\ref{entropy_v}), $\psi^{-1}$ denotes the inverse function of 
$\psi$ given by (\ref{psi}). Furthermore, 
$h_{f_{l}(w')\rightarrow l}^{\mathrm{dem}\rightarrow\mathrm{dec}}$ denotes the entropy 
sent from the demapper in section $f_{l}(w')$, given by (\ref{index_function}), 
to the variable nodes in section~$l$. 

We next analyze the check-to-variable message update. 
Assume that the check nodes in section~$l$ have received the entropy 
$h_{l-w\rightarrow l}^{\mathrm{v}\rightarrow\mathrm{c}}$ from the variable nodes in 
section~$l-w$ for $w\in[0:d_{\mathrm{v}})$. 
We approximate the distribution of the LLRs emitted from the variable 
nodes by $\mathcal{N}(\psi^{-1}(h_{l-w\rightarrow l}^{\mathrm{v}\rightarrow\mathrm{c}}),
2\psi^{-1}(h_{l-w\rightarrow l}^{\mathrm{v}\rightarrow\mathrm{c}}))$.  
In order to calculate the entropy $h_{l\rightarrow l-w}^{\mathrm{c}\rightarrow\mathrm{v}}$, 
we use the duality between variable nodes with entropy~$h$ and 
check nodes with entropy~$(1-h)$~\cite{Richardson08}. 
Exchanging the roles of variable nodes and check nodes, and repeating 
the derivation of (\ref{entropy_v}), we obtain  
\begin{IEEEeqnarray}{rl} 
h_{l\rightarrow l-w}^{\mathrm{c}\rightarrow\mathrm{v}} 
=& 1 - \psi\Biggl(
\left(
 \frac{d_{\mathrm{c}}}{d_{\mathrm{v}}} - 1 
\right)\psi^{-1}(1-h_{l-w\rightarrow l}^{\mathrm{v}\rightarrow\mathrm{c}}) 
\nonumber \\ 
+& \left. 
 \frac{d_{\mathrm{c}}}{d_{\mathrm{v}}}\sum_{w'\in[0:d_{\mathrm{v}}), w'\neq w}
 \psi^{-1}(1-h_{l-w'\rightarrow l}^{\mathrm{v}\rightarrow\mathrm{c}})
\right), \label{entropy_c} 
\end{IEEEeqnarray} 
with $h_{l-w'\rightarrow l}^{\mathrm{v}\rightarrow\mathrm{c}}=0$ for $l-w'<0$. 

Finally, we analyze the entropy emitted from the decoder in section~$l$. 
Suppose that the variable nodes in section~$l$ 
have received the entropy $h_{l+w\rightarrow l}^{\mathrm{c}\rightarrow\mathrm{v}}$ from 
the check nodes in section~$l+w$ for $w\in[0:d_{\mathrm{v}})$ in the last inner 
iteration. The entropy $h_{l}^{\mathrm{dec}\rightarrow\mathrm{dem}}$ 
for the LLRs emitted from the decoder in section~$l$ is given by 
\begin{equation} \label{output_entropy}
h_{l}^{\mathrm{dec}\rightarrow\mathrm{dem}} = 
\psi\left(
 \sum_{w=0}^{d_{\mathrm{v}}-1}\psi^{-1}(h_{l+w\rightarrow l}^{\mathrm{c}\rightarrow\mathrm{v}}) 
\right). 
\end{equation}
Similarly, the entropy $h_{l}^{\mathrm{dec}}$ of the a posteriori LLRs in 
section~$l$ is given by 
\begin{IEEEeqnarray}{r}  
h_{l}^{\mathrm{dec}} 
= \frac{1}{2W+1}\sum_{w=-W}^{W}\psi\Biggl(
 \psi^{-1}(h_{f_{l}(w)\rightarrow l}^{\mathrm{dem}\rightarrow\mathrm{dec}}) 
\nonumber \\
+\left.
 \sum_{w'=0}^{d_{\mathrm{v}}-1}
 \psi^{-1}(h_{l+w'\rightarrow l}^{\mathrm{c}\rightarrow\mathrm{v}})
\right), \label{decision_entropy}
\end{IEEEeqnarray}
which is associated with the decoding performance. 

The symmetry of linear codes implies that the entropy~(\ref{output_entropy}) 
is independent of the realizations of codewords. Thus, 
the pdf $p_{l}^{\mathrm{dec}}(L)$ of the LLRs emitted from the decoder in 
section~$l$ is approximated by the mixture Gaussian pdf
\begin{equation} \label{LLR_distribution} 
p_{l}^{\mathrm{dec}}(L) 
= \frac{1}{2}\sum_{a=\pm1}p_{\mathrm{G}}(L;am_{l}^{\mathrm{dec}\rightarrow\mathrm{dem}},
2m_{l}^{\mathrm{dec}\rightarrow\mathrm{dem}}), 
\end{equation} 
with $m_{l}^{\mathrm{dec}\rightarrow\mathrm{dem}}=
\psi^{-1}(h_{l}^{\mathrm{dec}\rightarrow\mathrm{dem}})$. 
In (\ref{LLR_distribution}), $p_{\mathrm{G}}(\cdot:m,\sigma^{2})$ denotes   
the real Gaussian pdf with mean $m$ and variance~$\sigma^{2}$.

We summarize the DE analysis for the inner iteration in stage~$l'$ of the SW 
decoding with the on-demand check node updating schedule. 
\begin{enumerate}
\item Let $j=1$. For all $l\in[l':l'+W_{\mathrm{SW}}+d_{\mathrm{v}}-2]$ and 
$w\in[0:d_{\mathrm{v}})$, let $h_{l\rightarrow l-w}^{\mathrm{c}\rightarrow\mathrm{v}}=1$ if 
$h_{l\rightarrow l-w}^{\mathrm{c}\rightarrow\mathrm{v}}$ is not initialized. Otherwise, 
use the current value. For all $l\in[l':l'+W_{\mathrm{SW}})$ and 
$w\in[0:d_{\mathrm{v}})$, update the entropy 
$h_{l\rightarrow l+w}^{\mathrm{v}\rightarrow\mathrm{c}}$ with (\ref{entropy_v}). 
\item Repeat the following in the order 
$l=l',\ldots,l'+W_{\mathrm{SW}}-1$:
\begin{itemize} 
\item Update $h_{l+w\rightarrow l}^{\mathrm{c}\rightarrow\mathrm{v}}$ with 
(\ref{entropy_c}) for all $w\in[0:d_{\mathrm{v}})$.
\item Update $h_{l\rightarrow l+w}^{\mathrm{v}\rightarrow\mathrm{c}}$ with 
(\ref{entropy_v}) for all $w\in[0:d_{\mathrm{v}})$. 
\end{itemize}
\item Let $j:=j+1$ and go back to Step 2) if $j$ is smaller than the total 
number $J$ of inner iterations. Otherwise, output the entropy 
$h_{l}^{\mathrm{dec}\rightarrow\mathrm{dem}}$ given by (\ref{output_entropy}) for all 
$l\in[l':l'+W_{\mathrm{SW}})$.   
\end{enumerate} 

\subsection{Density Evolution for Outer Iteration} 
The DE analysis for the outer iteration is summarized for the SW schedule. 
\begin{itemize}
\item Initialize $h_{l}^{\mathrm{dec}\rightarrow\mathrm{dem}}=0$ for $l\in[-W:0)$ and 
$h_{l}^{\mathrm{dec}\rightarrow\mathrm{dem}}=1$ for $l\in[0:L)$. 
\item Repeat the following in the order $l'=0,\ldots,L-W_{\mathrm{SW}}$: 
\begin{enumerate}
\item Let $i=1$. 
\item For all $l\in[l'-W:l'+W_{\mathrm{SW}}+W)$, 
let $\hat{x}_{\mathrm{dec}}^{2}(l)=0$ if $h_{l}^{\mathrm{dec}\rightarrow\mathrm{dem}}=1$. 
Otherwise, 
evaluate $\hat{x}_{\mathrm{dec}}^{2}(l)$ given by (\ref{squared_decision}) 
via the LLR distribution~(\ref{LLR_distribution}).  
\item Calculate the asymptotic MSE~(\ref{asymptotic_MSE}) based on 
Proposition~\ref{proposition1}, and then solve the FP equation~(\ref{FP}) 
in Proposition~\ref{proposition2} with the LLR 
distribution~(\ref{LLR_distribution}) for all $l\in[l'-W:l'+W_{\mathrm{SW}}+W)$. 
\item Estimate the average entropy 
$h_{l\rightarrow f_{l}(w)}^{\mathrm{dem}\rightarrow\mathrm{dec}}$ 
for all $l\in[l'-W:l'+W_{\mathrm{SW}}+W)$ and $w\in[-W:W]$, via numerical 
sampling of the LLRs~(\ref{ext_LLR_AWGN}) based on the LLR 
distributions~(\ref{LLR_distribution}) and the AWGN channel~(\ref{AWGN}) with 
SNR $(1-\xi_{l})/\sigma_{\mathrm{dem}}^{2}(l)$. 
\item Calculate the entropy $h_{l}^{\mathrm{dec}\rightarrow\mathrm{dem}}$ given by 
(\ref{output_entropy}) 
for all $l\in[l':l'+W_{\mathrm{SW}})$, by performing the DE analysis for the 
inner iteration, presented in Section~\ref{sec4-E}. 
\item Let $i:=i+1$ and go back to Step~2) if $i$ is smaller than the total 
number~$I$ of outer iterations. Otherwise, output the entropy 
$h_{l'}^{\mathrm{dec}}$ given by (\ref{decision_entropy}). 
\end{enumerate}
\end{itemize}

We shall define the threshold for the iterative CE-MUDD.  
When the window size $W_{\mathrm{SW}}$ tends to infinity, there exists some 
threshold $\rho>0$ such that the entropy $h_{l}^{\mathrm{dec}}$ of the a 
posteriori LLRs converges to zero for all SNRs $1/N_{0}>\rho$, whereas 
$h_{l}^{\mathrm{dec}}$ is strictly positive in some section $l$ for all 
$1/N_{0}<\rho$. However, the entropy $h_{l}^{\mathrm{dec}}$ may not converge to 
zero~\cite{Lentmaier102}, as long as the window size $W_{\mathrm{SW}}$ is finite. 
In other words, the average bit error ratio (BER) may not tend to zero. 
In order to present a formal definition of the threshold for the SW schedule, 
we define the average BER in section~$l$ as 
\begin{equation} \label{BER} 
p_{\mathrm{BER}}(h_{l}^{\mathrm{dec}}) 
= \mathrm{Q}\left(
 \sqrt{
  \frac{\psi^{-1}(h_{l}^{\mathrm{dec}})}{2}
 }
\right), 
\end{equation}
with $\mathrm{Q}(\cdot)$ denoting the Q-function. In (\ref{BER}), the 
entropy $h_{l}^{\mathrm{dec}}$ is given by (\ref{decision_entropy}). 

\begin{definition} \label{definition2} 
For given $\epsilon\geq 0$, the threshold $\rho_{\mathrm{BP}}$ for the iterative 
CE-MUDD is defined as the infimum of the SNR~$\rho$ such that, 
after $I$ outer iterations, the average BERs 
$p_{\mathrm{BER}}(h_{l}^{\mathrm{dec}})$ converge to values below $\epsilon$ 
for all $1/N_{0}>\rho$ and $l\in[0:L)$. 
\end{definition}

The iterative CE-MUDD can achieve an average BER below $\epsilon$ 
in the large-system limit after taking $M\to\infty$ if and only if 
the SNR $1/N_{0}$ is larger than the threshold. When the window size 
$W_{\mathrm{SW}}$ tends to infinity, the iterative CE-MUDD can achieve zero 
average BER for all SNRs $1/N_{0}$ greater than a threshold 
for $\epsilon=0$. When the window size $W_{\mathrm{SW}}$ is finite, 
on the other hand, the average BER achieved by the iterative CE-MUDD decreases 
toward zero very quickly, as the SNR grows from a threshold for 
strictly positive $\epsilon>0$. However, it is open whether 
there exists a finite threshold $\rho_{\mathrm{BP}}$ such that zero 
average BER is achieved for all $1/N_{0}>\rho_{\mathrm{BP}}$ when 
a finite window size $W_{\mathrm{SW}}$ is used. 

\section{Numerical Results} \label{section5} 
\subsection{Density Evolution}
In all numerical results, we consider $6\times6$ MIMO systems over 
Rayleigh block-fading channels with coherence time~$T=64$. 
As noted in Remark~\ref{remark1}, the noise variance $\tilde{N}_{0}$ 
postulated in the channel estimator is assumed to be equal to the true one 
$N_{0}$ in the DE analysis, whereas a mismatched value $\tilde{N}_{0}\neq N_{0}$ 
may be used in numerical simulations. 

\begin{table}[t]
\begin{center}
\caption{
Four systems compared in this paper.
}
\label{table_system}
\begin{tabular}{|c||c|c|}
\hline
& coding & BICM \\ 
\hline\hline 
Conventional system & LDPC & $W=0$ \\ 
\hline
Coupled system in BICM & LDPC & $W\geq1$ \\ 
\hline
Coupled system in coding & SC LDPC & $W=0$ \\
\hline 
Coupled system in both coding and BICM & SC LDPC & $W\geq1$ \\
\hline 
\end{tabular}
\end{center}
\end{table}

We compare four systems in terms of the decoding threshold, shown in 
Table~\ref{table_system}. 
One system consists of $(3, 6)$ LDPC coding and conventional BICM with $W=0$, 
and is called a conventional system. The BP algorithm is assumed in decoding. 
In a second system, the STC BICM with $W\geq 1$ is used instead of the 
conventional BICM. Since the performance is poor in sections at the right end, 
we consider STC BICM with $2(L+W)$ sections in which known words are sent in 
sections $l$ at both ends for $l\in[-W:0)$ and $l\in[2L:2L+W)$. 
In the SW decoding, two decoding windows run from both ends toward the center. 
The system corresponds to those in \cite{Takeuchi134,Horio15}, and has the 
same overall rate as (\ref{rate}) for the other systems. 
We refer to this STC BICM as $(M, L, W, Q)$ both-side STC BICM 
or simply as $(M, L, W, Q)$ STC BICM. 

A third system is constructed from $(3, 6, L)$ SC LDPC 
coding~(\ref{base_matrix}) with efficient 
termination for encoding~\cite{Tazoe12} and from the conventional BICM with 
$W=0$. The third system is called the coupled system in coding, whereas 
the second system is referred to as the coupled system in BICM. 
The third system was investigated for the AWGN channel in \cite{Schmalen13}. 
In the last system, coupling is introduced for both coding and BICM. 
The system consists of $(3, 6, L)$ SC LDPC coding~(\ref{base_matrix}) and 
$(M, L, W, Q)$ one-side STC BICM. Thus, we referred to the last system as 
the coupled system in both coding and BICM. 

\begin{figure}[t]
\begin{center}
\includegraphics[width=\hsize]{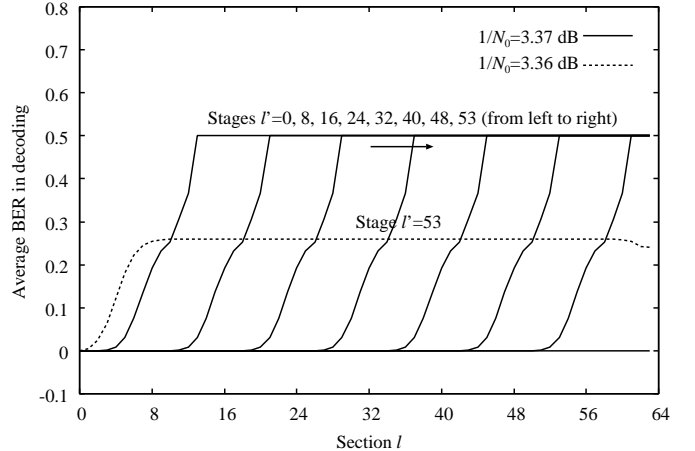}
\end{center}
\caption{
Average BER in decoding versus section~$l$ for $(3, 6, 64)$ SC LDPC coding,  
$(\infty, 64, 1, 2)$ STC BICM with QPSK, the SW schedule with 
window size~$W_{\mathrm{SW}}=11$ and the numbers~$J=\infty$ and $I=\infty$ of 
inner and outer iterations, and $6\times6$ block-fading MIMO with 
coherence time~$T=64$ and the number of pilots~$T_{\mathrm{tr}}=6$. 
}
\label{fig5}
\end{figure}

We first present the dynamics of the SW decoding as the decoding stage 
proceeds. Figure~\ref{fig5} shows the average BER in decoding versus 
section~$l$ for the coupled system in both coding and BICM with QPSK. 
The decoding proceeds from left to right, and eventually almost zero average 
BER is accomplished for $1/N_{0}=3.37$~dB, whereas the average BER is distinct 
from zero for $1/N_{0}=3.36$~dB. 

\begin{figure}[t]
\begin{center}
\includegraphics[width=\hsize]{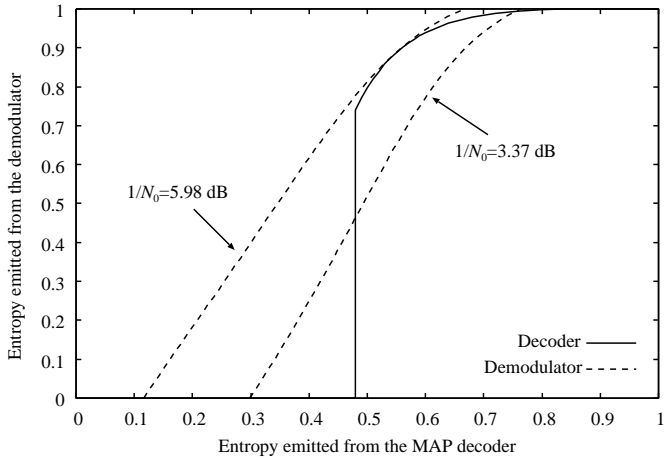}
\end{center}
\caption{
EXIT chart for $(3, 6)$ LDPC coding under the MAP decoding, conventional BICM 
with QPSK ($W=0$), and $6\times6$ block-fading MIMO with coherence time~$T=64$ 
and the number of pilots~$T_{\mathrm{tr}}=6$. 
}
\label{fig6}
\end{figure}

Figure~\ref{fig6} shows the extrinsic information transfer (EXIT) chart for 
the underlying $(3, 6)$ LDPC coding and the conventional BICM with QPSK. 
The SNR $1/N_{0}=5.98$~dB is approximately equal to the threshold for the 
combination of the conventional BICM and the $(3, 6)$ LDPC coding under the 
MAP decoding, since the two EXIT curves have the unique intersection at 
$(0.116, 0)$. Thus, the threshold for the conventional system would be worse 
than $5.98$~dB, since the BP decoding is actually used instead of the MAP 
decoding. As shown in Fig.~\ref{fig5}, on the other hand, the SNR 
$1/N_{0}=3.37$~dB is achievable by the coupled system in both coding and BICM 
under the SW decoding. Thus, coupling can provide a performance gain of 
$2.61$~dB. We observe that the two EXIT curves have three points of 
intersection at the SNR $1/N_{0}=3.37$~dB. If coupling were not used, 
the system would converge to the top FP that has the maximum entropy 
among the three FPs. Coupling allows the system to converge toward 
the bottom FP with zero entropy even when there are multiple FPs. 

\begin{table}[t]
\begin{center}
\caption{
Thresholds for $(3, 6)$ LDPC (top) and $(3, 6, \infty)$ SC LDPC (bottom) 
coding. We used $(\infty, \infty, W, 2)$ STC BICM with QPSK, 
the SW schedule with infinite window size~$W_{\mathrm{SW}}=\infty$ and the 
number of outer iterations~$I=\infty$, 
the BER requirement $\epsilon=0$, 
and $6\times6$ block-fading MIMO with coherence time~$T=64$. 
}
\label{table1}
\begin{tabular}{|c||c|c|c|c|c|}
\hline
& $T_{\mathrm{tr}}=0$ & $T_{\mathrm{tr}}=2$ & $T_{\mathrm{tr}}=4$ & $T_{\mathrm{tr}}=6$ 
& Perfect CSI \\ 
\hline\hline
& $\infty$~dB & $17.3$~dB & $7.40$~dB & $5.98$~dB & $2.94$~dB \\ 
\cline{2-6} 
$W=0$ & $\infty$~dB & $3.76$~dB & $3.54$~dB & $3.37$~dB & $1.69$~dB \\
\hline\hline
& $5.39$~dB & $4.87$~dB & $4.50$~dB & $4.24$~dB & $2.36$~dB \\ 
\cline{2-6} 
$W=1$ & $4.04$~dB & $3.76$~dB & $3.54$~dB & $3.37$~dB & $1.67$~dB \\ 
\hline\hline
& $5.04$~dB & $4.62$~dB & $4.31$~dB & $4.07$~dB & $2.28$~dB \\ 
\cline{2-6} 
$W=2$ & $4.04$~dB & $3.76$~dB & $3.54$~dB & $3.36$~dB & $1.66$~dB \\
\hline
\end{tabular}
\end{center}
\end{table}
\begin{table}[t]
\begin{center}
\caption{
Thresholds for $(3, 6)$ LDPC (top) and $(3, 6, \infty)$ SC LDPC (bottom) 
coding. The other conditions are the same as in Table~\ref{table1}, 
with the exception of $(\infty, \infty, W, 4)$ STC BICM with $16$ QAM. 
}
\label{table2}
\begin{tabular}{|c||c|c|c|c|c|}
\hline
& $T_{\mathrm{tr}}=0$ & $T_{\mathrm{tr}}=2$ & $T_{\mathrm{tr}}=4$ & $T_{\mathrm{tr}}=6$ 
& Perfect CSI \\ 
\hline\hline
& $\infty$~dB & $\infty$~dB & $\infty$~dB & $16.3$~dB & $10.8$~dB \\ 
\cline{2-6} 
$W=0$ & $\infty$~dB & $13.2$~dB & $11.9$~dB & $11.1$~dB & $8.2$~dB \\
\hline\hline
& $19.5$~dB & $15.2$~dB & $13.4$~dB & $12.2$~dB & $8.8$~dB \\ 
\cline{2-6} 
$W=1$ & $13.9$~dB & $12.5$~dB & $11.5$~dB & $10.8$~dB & $8.0$~dB \\ 
\hline\hline
& $16.3$~dB & $13.8$~dB & $12.4$~dB & $11.6$~dB & $8.6$~dB \\ 
\cline{2-6} 
$W=2$ & $13.3$~dB & $12.1$~dB & $11.2$~dB & $10.6$~dB & $7.9$~dB \\
\hline
\end{tabular}
\end{center}
\end{table}

Table~\ref{table1} lists the thresholds of the iterative CE-MUDD for the 
four systems with QPSK. Note that the overall rate~(\ref{rate}) is 
equal to $R=(1-T_{\mathrm{tr}}/T)QKr$, with the design rate $r=1/2$ in coding, 
for all systems, since the limit $L\rightarrow\infty$ is assumed. 
Thus, each column in the table contains systems with the same rate $R$. 
We simulated the numbers of inner iterations $J=1$ and $J=\infty$, and 
found that the thresholds are the same as each other for the two cases. 
Thus, the thresholds are independent of the number of inner iterations $J$,  
as long as the number of outer iterations $I=\infty$ is considered. 
From Definition~\ref{definition2}, the decoder can achieve the 
BER requirement $\epsilon=0$ for all sections if and only if 
the SNR $1/N_{0}$ is larger than the corresponding threshold. Infinite 
thresholds imply that the average BER cannot tend to zero for $N_{0}=0$. 
We find that the thresholds improve for all cases when coupling is introduced 
for coding. Furthermore, the coupled system in coding outperforms the coupled 
system in BICM~\cite{Takeuchi134,Horio15}, with the exception of 
$T_{\mathrm{tr}}=0$. On the other hand, the thresholds hardly improve 
as the coupling width~$W$ in BICM increases, as long as SC LDPC coding is 
used. We conclude that, for QPSK, it is sufficient to introduce coupling 
only in coding.  

\begin{table}[t]
\begin{center}
\caption{
Thresholds for $(3, 6)$ LDPC (top) and $(3, 6, \infty)$ SC LDPC (bottom) 
coding. The other conditions are the same as in Table~\ref{table1}, 
with the exception of $(\infty, \infty, W, 6)$ STC BICM with $64$ QAM. 
}
\label{table3}
\begin{tabular}{|c||c|c|c|c|c|}
\hline
& $T_{\mathrm{tr}}=4$ & $T_{\mathrm{tr}}=8$ & $T_{\mathrm{tr}}=12$ & $T_{\mathrm{tr}}=16$ 
& Perfect CSI \\ 
\hline\hline
& $\infty$~dB & $23.8$~dB & $21.2$~dB & $20.3$~dB & $18.3$~dB \\ 
\cline{2-6} 
$W=0$ & $25.6$~dB & $17.7$~dB & $16.5$~dB & $16.0$~dB & $14.4$~dB \\
\hline\hline
& $26.3$~dB & $18.0$~dB & $16.7$~dB & $16.1$~dB & $14.4$~dB \\ 
\cline{2-6} 
$W=1$ & $21.4$~dB & $17.1$~dB & $16.0$~dB & $15.5$~dB & $13.9$~dB \\ 
\hline\hline
& $22.7$~dB & $17.3$~dB & $16.1$~dB & $15.6$~dB & $14.0$~dB \\ 
\cline{2-6} 
$W=2$ & $20.2$~dB & $16.5$~dB & $15.5$~dB & $15.0$~dB & $13.5$~dB \\
\hline
\end{tabular}
\end{center}
\end{table}

\begin{figure}[t]
\begin{center}
\includegraphics[width=\hsize]{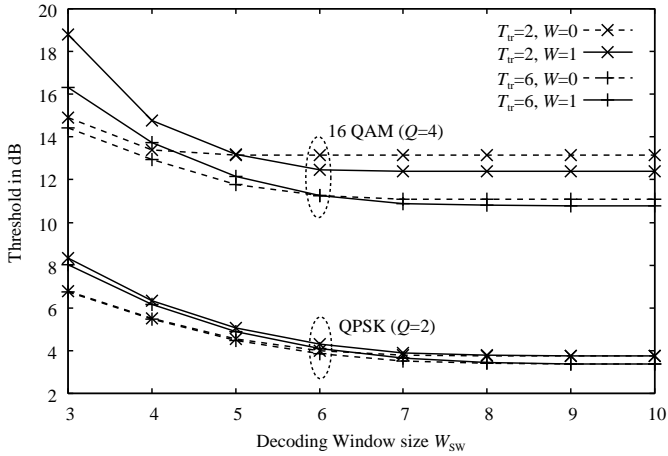}
\end{center}
\caption{
Threshold versus window size~$W_{\mathrm{SW}}$ for $(3, 6, 64)$ SC LDPC coding, 
$(\infty, 64, W, Q)$ STC BICM, the SW schedule with the numbers $J=\infty$ 
and $I=\infty$ of inner and outer iterations,  
the BER requirement $\epsilon=10^{-6}$, 
and $6\times6$ block-fading MIMO with coherence time~$T=64$. 
}
\label{fig7}
\end{figure}

The thresholds for $16$~QAM and $64$~QAM are shown in Tables~\ref{table2} 
and \ref{table3}. We observe that coupling in BICM can provide a significant 
improvement in the threshold especially for $64$~QAM, as well as coupling in 
coding. As a result, we find that the coupled system in BICM can outperform 
the coupled system in coding for $64$~QAM, by making comparisons between 
the top row for $W=2$ and the bottom row for $W=0$ in Table~\ref{table3}. 
We shall explain why coupling should be introduced for BICM, as well as for 
coding. The coupled system in BICM can utilize the data symbols decoded in 
the preceding stages as training symbols in the current stage. 
The training symbols reduce inter-stream interference in MUD and allow the 
receiver to obtain reliable initial channel estimates. Eventually, the 
coupled system in BICM can attain reliable decoding results in the current 
stage. On the other hand, the coupled system only in coding utilizes 
decoding results in the preceding stages only for decoding in the current 
stage. For QPSK, the receiver may attain reasonably good channel estimates 
and detection results in the initial outer iteration. However, 
it may not obtain them in the initial outer iteration for higher-order 
modulation. As a result, the coupled system in coding has worse threshold 
than the coupled system in both coding and BICM or only in BICM.   

We next consider how to select design parameters such as the window 
size $W_{\mathrm{SW}}$ and the number $I$ of outer iterations. 
Figure~\ref{fig7} shows the thresholds versus the window size $W_{\mathrm{SW}}$ 
for $(3, 6, 64)$ SC LDPC coding. We find that the thresholds converge for 
$W_{\mathrm{SW}}\geq9$, and that the thresholds for $W=0$ converges slightly 
more quickly than those for $W=1$. Thus, an option is $W_{\mathrm{SW}}=9$ for 
achieving the ultimate threshold based on $W_{\mathrm{SW}}=\infty$ 
approximately.  

\begin{figure}[t]
\begin{center}
\includegraphics[width=\hsize]{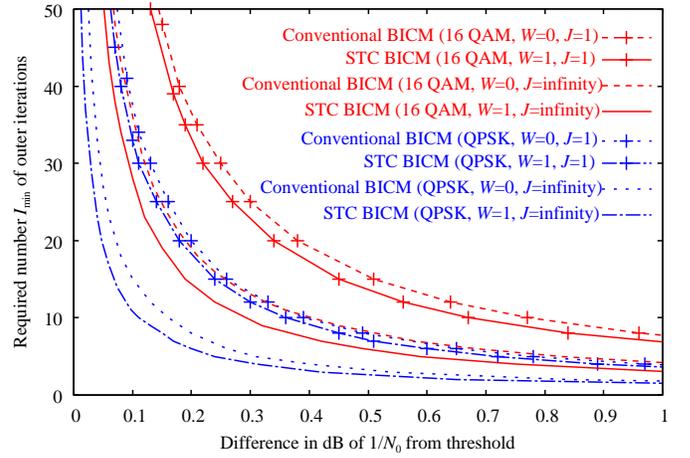}
\end{center}
\caption{
Required number $I_{\mathrm{min}}$ of outer iterations versus the difference 
$1/N_{0}-\rho_{\mathrm{BP}}$ of the SNR from the threshold $\rho_{\mathrm{BP}}$ 
for $(3, 6, 64)$ SC LDPC coding, $(\infty, 64, W, Q)$ STC BICM, 
the SW schedule with the window size $W_{\mathrm{SW}}=9$, 
the BER requirement $\epsilon=10^{-6}$, 
and $6\times6$ block-fading MIMO with coherence time~$T=64$ 
and the number of pilots $T_{\mathrm{tr}}=6$. 
}
\label{fig8}
\end{figure}

We investigate the minimum $I_{\mathrm{min}}$ of the number of outer iterations 
such that the BER requirement $\epsilon$ is satisfied in all sections for 
an SNR~$1/N_{0}>\rho_{\mathrm{BP}}$ above the threshold $\rho_{\mathrm{BP}}$ 
for infinite outer iterations $I=\infty$. In order to eliminate boundary 
effect, we assume infinite outer iterations for the first and last stages of 
the SW decoding. As shown in Fig.~\ref{fig8}, the required number 
$I_{\mathrm{min}}$ in the bulk region reduces quickly as the SNR $1/N_{0}$ 
increases from the ultimate threshold $\rho_{\mathrm{BP}}$ for $I=\infty$. 
We observe that the STC BICM with 
$W=1$ results in smaller $I_{\mathrm{min}}$ than the conventional BICM with 
$W=0$. The required numbers $I_{\mathrm{min}}$ for the numbers of inner 
iterations $J=1$ and $J=\infty$ correspond to upper and lower bounds on 
the required number for a finite number of inner iterations, respectively. 
For example, the BER requirement is achieved by selecting the number of outer 
iterations $I\in[8:20]$ for $16$~QAM and $W=1$, when one tolerates 
an SNR loss of $0.35$~dB from the threshold $\rho_{\mathrm{BP}}$ for $I=\infty$. 

\subsection{Numerical Simulations}
We have so far considered the infinite code-length limit~$M$. Next, 
numerical simulations for finite $M$ are presented. For the coupled systems, 
we assumed the code length $M=4QK(T-T_{\mathrm{tr}})$ such that 
the frame length including pilot symbols was equal to the length of $4$ 
fading blocks in each section. For the conventional system, 
on the other hand, we assumed $M=4W_{\mathrm{SW}}QK(T-T_{\mathrm{tr}})$ with 
$W_{\mathrm{SW}}$ denoting the window size for the SW decoding in the coupled 
systems.  

\begin{figure}[t]
\begin{center}
\includegraphics[width=\hsize]{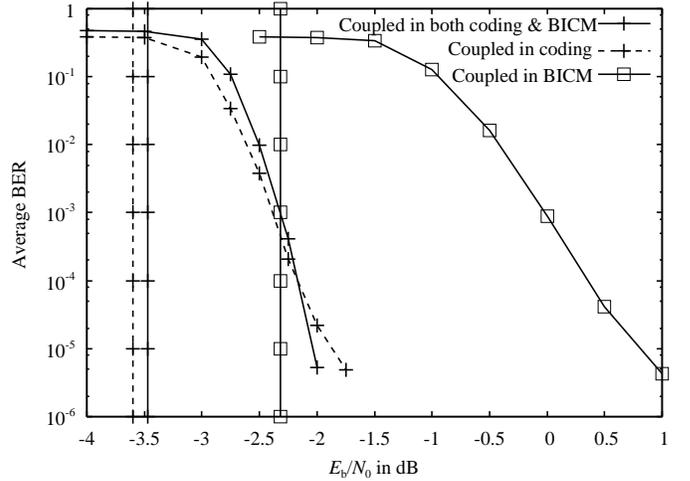}
\end{center}
\caption{
Average BER versus $E_{\mathrm{b}}/N_{0}$ for QPSK, 
the SW decoding with the window size $W_{\mathrm{SW}}=9$ and the numbers of inner 
and outer iterations~$J=4$ and $I=20$, and $6\times6$ block-fading MIMO 
with coherence time~$T=64$. $\tilde{N}_{0}=N_{0}$ was used. 
(i) Coupled system in BICM: $(3, 6)$ LDPC coding, $(3024, 62, 1, 2)$ STC BICM, 
and $T_{\mathrm{tr}}=1$. (ii) Coupled system in coding: 
$(3, 6, 63)$ SC LDPC coding, $(3024, 63, 0, 2)$ conventional BICM, 
and $T_{\mathrm{tr}}=1$. 
(iii) Coupled system in both coding and BICM: $(3, 6, 63)$ SC LDPC coding, 
$(3072, 63, 1, 2)$ STC BICM, and $T_{\mathrm{tr}}=0$. The vertical lines 
represent the corresponding thresholds for the BER requirement 
$\epsilon=10^{-6}$. 
}
\label{fig9}
\end{figure}

Figure~\ref{fig9} displays the average BERs in decoding versus 
$E_{\mathrm{b}}/N_{0}=1/(RN_{0})$ for QPSK. 
The overall rate~$R$ given by (\ref{rate}) is equal to $R=5.8125$~bps/Hz 
for all systems. We omitted the conventional system that has $(3, 6)$ LDPC 
coding, $(9\times2976, 62, 0, 2)$ conventional BICM, and $T_{\mathrm{tr}}=2$, 
since the ultimate threshold $E_{\mathrm{b}}/N_{0}\approx9.7$~dB 
is too bad, as indicated from Table~\ref{table1}. 
We find that the thresholds for the coupled systems in coding provide good 
predictions for the locations of the so-called waterfall regime, whereas 
the threshold under-estimates the location for the coupled system in BICM. 
One interesting observation is that the coupled system in both coding and 
BICM has a steeper BER slope in the waterfall regime than that only in coding. 
Consequently, the coupled system in both coding and BICM can achieve the 
best performance in the high SNR regime, even though the best threshold is 
achieved by the coupled system in coding. 

\begin{figure}[t]
\begin{center}
\includegraphics[width=\hsize]{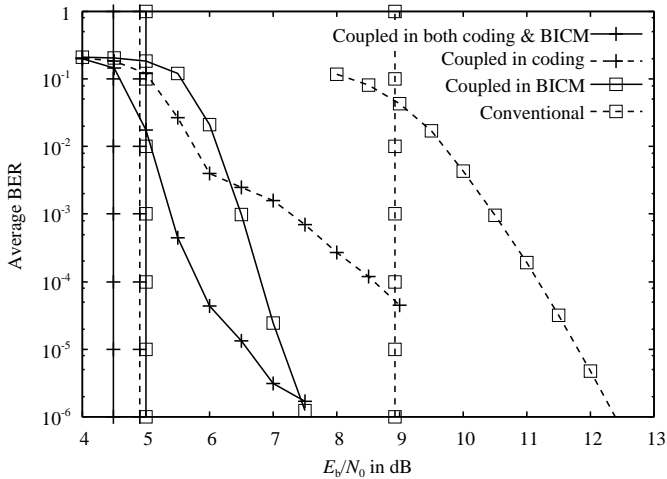}
\end{center}
\caption{
Average BER versus $E_{\mathrm{b}}/N_{0}$ for $64$ QAM and $6\times6$ 
block-fading MIMO with coherence time~$T=64$. 
$1/\tilde{N}_{0}=14$~dB ($E_{\mathrm{b}}/\tilde{N}_{0}\approx2.89$~dB) was used. 
(i) Conventional system: $(3, 6)$ LDPC coding, $(9\times6624, 1, 0, 6)$ 
conventional BICM, and $T_{\mathrm{tr}}=18$. 
(ii) Coupled system in BICM: $(3, 6)$ LDPC coding, $(6768, 46, 1, 6)$ STC BICM, 
and $T_{\mathrm{tr}}=17$. (iii) Coupled system in coding: $(3, 6, 47)$ SC LDPC 
coding, $(6768, 47, 0, 6)$ conventional BICM, and $T_{\mathrm{tr}}=17$. 
(iv) Coupled system in both coding and BICM: $(3, 6, 47)$ SC LDPC coding, 
$(6912, 47, 1, 6)$ STC BICM, and $T_{\mathrm{tr}}=16$. The coupled systems use 
the SW decoding with the window size $W_{\mathrm{SW}}=9$ and the numbers of inner 
and outer iterations~$J=4$ and $I=20$, whereas $J=12$ and $I=60$ are 
used for the conventional system. The vertical lines 
represent the corresponding thresholds for the BER requirement 
$\epsilon=10^{-6}$. 
}
\label{fig10}
\end{figure}

Figure~\ref{fig10} shows numerical simulations for $64$ QAM. 
The overall rate $R$ is equal to $R=12.9375$~bps/Hz for all systems. 
The error floor occurs remarkably for the coupled system in coding, even 
though the noise variance $\tilde{N}_{0}$ postulated in the channel estimator 
was selected so as to reduce this error floor. As a result, the two systems 
coupled in BICM outperform the coupled system in coding for high SNRs.  

The error floor for the coupled system in coding is because the SW decoding 
fails to propagate in an intermediate section. In the initial outer iteration, 
the channel estimator has to estimate channel gains only from pilot symbols 
at the rightmost section of each decoding window, whereas it can utilize 
feedback information from the decoder in the other sections. Consequently, 
messages at the rightmost section are unreliable in the initial outer 
iteration. For finite-sized systems, the unreliable messages may propagate 
from right to left with a small probability. When such an unexpected 
propagation occurs in a stage of the SW decoding, reliable information 
fails to propagate from left to right.  

The STC BICM can improve reliability of initial messages at the rightmost 
section in each decoding window, since a portion of the preceding decoding 
results can be utilized for the initial channel estimator at the rightmost 
section. This improvement results in a significant reduction of the average 
BER in the high SNR regime for the coupled system in both coding and BICM, 
as well as in an improvement of the decoding threshold.  

\section{Conclusion} \label{section6} 
We have considered coupling in coding and BICM to improve the performance 
of iterative CE-MUDD. Coupling in coding results in a dominant improvement 
in the performance for low rate systems with QPSK. Coupling in BICM can 
provide additional significant gains in the waterfall and high SNR regimes 
for higher rate systems with higher-order modulation. We conclude that 
spatial coupling should be introduced for BICM in MIMO systems with 
higher-order modulation. 

We could not conclude that spatial coupling should be introduced for 
{\em both} coding and BICM for higher-order modulation, since an error 
floor occurs when spatial coupling is used in coding. The analysis of this 
error floor is left as a future work. 

\appendices

\section{Efficient Calculation of (\ref{posterior_covariance})}
\label{appendix_A} 
We shall present an efficient method for calculating 
(\ref{posterior_covariance}) for all $t$. 
Let us define 
\begin{equation} \label{Xi} 
\boldsymbol{\Xi}^{\mathrm{ch}} 
= \left\{
 K\boldsymbol{I}_{K} 
 + \hat{\boldsymbol{X}}(\boldsymbol{\Sigma}^{\mathrm{ch}}
 +\tilde{N}_{0}\boldsymbol{I}_{T})^{-1}\hat{\boldsymbol{X}}^{\mathrm{H}}
\right\}^{-1}, 
\end{equation}
with $\boldsymbol{\Sigma}^{\mathrm{ch}}
=\mathrm{diag}\{\sigma_{t}^{2}:\hbox{all $t$}\}$, 
in which $\sigma_{t}^{2}$ is given by (\ref{Sigma_x}). Using the matrix 
inversion lemma\footnote{
$(\boldsymbol{A}+\boldsymbol{B}\boldsymbol{D}\boldsymbol{C})^{-1}
=\boldsymbol{A}^{-1}-\boldsymbol{A}^{-1}\boldsymbol{B}
(\boldsymbol{D}^{-1}+\boldsymbol{C}\boldsymbol{A}^{-1}\boldsymbol{B})^{-1}
\boldsymbol{C}\boldsymbol{A}^{-1}$. 
} for the identity 
\begin{equation}
(\boldsymbol{\Xi}_{\backslash t}^{\mathrm{ch}})^{-1} 
= (\boldsymbol{\Xi}^{\mathrm{ch}})^{-1} 
- \frac{\hat{\boldsymbol{x}}_{t}\hat{\boldsymbol{x}}_{t}^{\mathrm{H}}}
{\sigma_{t}^{2} + \tilde{N}_{0}}, 
\end{equation}
with $\hat{\boldsymbol{x}}_{t}$ denoting the $t$th column of 
$\hat{\boldsymbol{X}}$, we obtain 
\begin{equation} \label{Xi_t} 
\boldsymbol{\Xi}_{\backslash t}^{\mathrm{ch}} 
= \boldsymbol{\Xi}^{\mathrm{ch}} 
- \frac{\boldsymbol{\Xi}^{\mathrm{ch}}\hat{\boldsymbol{x}}_{t}
(\boldsymbol{\Xi}^{\mathrm{ch}}\hat{\boldsymbol{x}}_{t})^{\mathrm{H}}}
{\hat{\boldsymbol{x}}_{t}^{\mathrm{H}}\boldsymbol{\Xi}^{\mathrm{ch}}
\hat{\boldsymbol{x}}_{t} - (\sigma_{t}^{2} + \tilde{N}_{0})}. 
\end{equation}
Expression~(\ref{Xi_t}) implies that all a posteriori covariance 
matrices~(\ref{posterior_covariance}) can be obtained by one calculation of 
the inverse matrix~(\ref{Xi}), instead of $T$ calculations. 


\section{Derivation of Proposition~\ref{proposition1}}
\label{proof_proposition1} 
\subsection{Formulation} 
The derivation of Proposition~\ref{proposition1} is an extension of the 
replica method in \cite{Takeuchi134}. Assume $\tilde{N}_{0}=N_{0}$. 
Since the received matrix 
$\boldsymbol{Y}_{\backslash t}$ given by (\ref{MIMO_t}) has i.i.d.\ rows, 
without loss of generality, we focus on the first row 
$\vec{\boldsymbol{y}}_{1,\backslash t}$ and drop the subscript~$1$ 
from all variables. For notational convenience, we write 
the first row of the channel matrix as 
$K^{-1/2}\vec{\boldsymbol{h}}_{0}\in\mathbb{C}^{1\times K}$, with 
$\vec{\boldsymbol{h}}_{0}\sim\mathcal{CN}(\boldsymbol{0},\boldsymbol{I}_{K})$. 
Let $\vec{\boldsymbol{h}}_{a}\in\mathbb{C}^{1\times K}$ denote the replicas 
of $\vec{\boldsymbol{h}}_{0}$: 
$\mathcal{H}=\{\vec{\boldsymbol{h}}_{a}:a=0,\ldots,n\}$ are independent CSCG 
vectors with covariance $\boldsymbol{I}_{K}$. Let us define a function 
$Z_{t}(n,\omega)$ as 
\begin{IEEEeqnarray}{rl} 
Z_{t}(n,\omega) 
=
& 
\mathbb{E}\left[
 \int e^{\omega f}\left\{
  \int 
  p(\vec{\boldsymbol{y}}_{\backslash t} | \vec{\boldsymbol{h}}_{1}, 
  \hat{\boldsymbol{X}}_{\backslash t})p(\vec{\boldsymbol{h}}_{1})
  d\vec{\boldsymbol{h}}_{1} 
 \right\}^{n-2} 
\right.
\nonumber \\ &
\left.
 \cdot
\prod_{a=0}^{2}\left\{
  p(\vec{\boldsymbol{y}}_{\backslash t} | 
  \vec{\boldsymbol{h}}_{a}, 
  \hat{\boldsymbol{X}}_{\backslash t})p(\vec{\boldsymbol{h}}_{a})
  d\vec{\boldsymbol{h}}_{a} 
 \right\}
 d\vec{\boldsymbol{y}}_{\backslash t} 
\right], \label{partition_function} 
\end{IEEEeqnarray}
with 
\begin{equation}
f(\vec{\boldsymbol{h}}_{0},\vec{\boldsymbol{h}}_{1},\vec{\boldsymbol{h}}_{2}) 
= \sum_{k=1}^{K}\{(\vec{\boldsymbol{h}}_{0})_{k} 
- (\vec{\boldsymbol{h}}_{1})_{k}\}\{(\vec{\boldsymbol{h}}_{0})_{k} 
- (\vec{\boldsymbol{h}}_{2})_{k}\}^{*}. 
\end{equation}
In (\ref{partition_function}), the pdf 
$p(\vec{\boldsymbol{y}}_{\backslash t} | \vec{\boldsymbol{h}}_{0}, 
\hat{\boldsymbol{X}}_{\backslash t})$ denotes the conditional pdf for the 
first row $\vec{\boldsymbol{y}}_{\backslash t}$ of the received matrix given by 
(\ref{MIMO_t}). 
Furthermore, $p(\vec{\boldsymbol{y}}_{\backslash t} | \vec{\boldsymbol{h}}_{a}, 
\hat{\boldsymbol{X}}_{\backslash t})$ for $a\geq1$ is the conditional pdf of 
$\vec{\boldsymbol{y}}_{\backslash t}$ after making the Gaussian approximation of 
the second term on the RHS of (\ref{MIMO_t}), given by
\begin{equation} \label{Gaussian_MIMO} 
p(\vec{\boldsymbol{y}}_{\backslash t} | \vec{\boldsymbol{h}}_{a}, 
\hat{\boldsymbol{X}}_{\backslash t}) 
= \prod_{t'\neq t}g_{\mathrm{CG}}\left(
 y_{t'};\frac{1}{\sqrt{K}}\vec{\boldsymbol{h}}_{a}\hat{\boldsymbol{x}}_{t'},
 \sigma_{t'}^{2} +N_{0}
\right), 
\end{equation}
where $g_{\mathrm{CG}}(\cdot;m,\sigma^{2})$ denotes 
the proper complex Gaussian pdf with mean $m$ and variance $\sigma^{2}$. 
In (\ref{Gaussian_MIMO}), $y_{t'}$ and $\hat{\boldsymbol{x}}_{t'}$ denote 
the $(1,t')$-element of $\boldsymbol{Y}$ and the $t'$th column of 
$\hat{\boldsymbol{X}}$, respectively. The variance $\sigma_{t'}^{2}$ is given 
by~(\ref{Sigma_x}).    

\begin{lemma}{\cite[Lemma~2]{Takeuchi134}} \label{lemma1} 
The average MSE~(\ref{average_MSE}) is given by  
\begin{equation} \label{target} 
\xi_{\backslash t}(l)=
\lim_{n\downarrow 0}\lim_{\omega\to0}\frac{1}{K}\frac{\partial}{\partial\omega}
\ln Z_{t}(n,\omega).  
\end{equation}
\end{lemma}  

Lemma~\ref{lemma1} implies that evaluating the average MSE~(\ref{average_MSE}) 
reduces to calculating the quantity~(\ref{target}) in the large-system limit. 
We follow \cite{Takeuchi134} to evaluate (\ref{target}) with the replica 
method. 

\subsection{Replica Method} 
We first calculate (\ref{partition_function}) only for integers $n\geq2$. 
The assumption of random interleaving implies that independent LLRs are 
fed back from the decoder for sufficiently long code length $M$. In other 
words, $x_{k,t'}$ and $\hat{x}_{k,t'}$ are independent for all $k$ and $t'$ 
in the limit $M\to\infty$. From the central limit theorem, the 
{\em individual} estimation error $K^{-1/2}\vec{\boldsymbol{h}}_{0}
(\boldsymbol{x}_{t'}-\hat{\boldsymbol{x}}_{t'})$ conditioned on 
$\vec{\boldsymbol{h}}_{0}$ and $\hat{\boldsymbol{x}}_{t'}$ converges 
in distribution to a CSCG vector with covariance $\sigma_{t'}^{2}$ 
given by (\ref{Sigma_x}) in the limit $K\to\infty$. 
Thus, each marginal pdf $p(y_{t'}|\vec{\boldsymbol{h}}_{0},
\hat{\boldsymbol{x}}_{t'})$ tends to the $t'$th factor on the RHS of 
(\ref{Gaussian_MIMO}) with $\vec{\boldsymbol{h}}_{a}
=\vec{\boldsymbol{h}}_{0}$.  
Note that we do not claim the convergence of the joint pdf  
$p(\vec{\boldsymbol{y}}_{\backslash t} | \vec{\boldsymbol{h}}_{0}, 
\hat{\boldsymbol{X}}_{\backslash t})$ to (\ref{Gaussian_MIMO}). From these 
observations, we find that (\ref{partition_function}) reduces to 
\begin{IEEEeqnarray}{l} 
\frac{1}{K}\ln Z_{t}(n,\omega) 
= \mathbb{E}\left[
 e^{\omega f}\left\{
  e_{n}(\{v_{a}^{\mathrm{p}}\},N_{0},\mathcal{H})  
 \right\}^{T_{\mathrm{tr}}}
\right. 
\nonumber \\ 
\left.
\cdot
 \prod_{t'\in(T_{\mathrm{tr}}:T], t'\neq t}
  e_{n}(\{v_{a}^{\mathrm{c}}\},\sigma_{t'}^{2}+N_{0},\mathcal{H}) 
\right] + O(K^{-1}) \label{partition2} 
\end{IEEEeqnarray}
in the large-system limit, with 
\begin{equation} \label{e} 
e_{n}(\{v_{a}\},\sigma^{2},\mathcal{H}) 
= \mathbb{E}\left[
 \left. 
  \int_{\mathbb{C}}\prod_{a=0}^{n}g_{\mathrm{CG}}(y;v_{a},\sigma^{2})dy
 \right| \mathcal{H} 
\right]. 
\end{equation} 
In (\ref{partition2}), $v_{a}^{\mathrm{p}}$ and $v_{a}^{\mathrm{c}}$ are 
respectively given by 
\begin{equation} 
v_{a}^{\mathrm{p}} 
= \frac{1}{\sqrt{K}}\sum_{k=1}^{K}(\vec{\boldsymbol{h}}_{a})_{k}x_{k,1},
\quad 
v_{a}^{\mathrm{c}} 
= \frac{1}{\sqrt{K}}\sum_{k=1}^{K}(\vec{\boldsymbol{h}}_{a})_{k}\hat{x}_{k,T}. 
\end{equation}

Recall that we are focusing on a fading block in section~$l$. 
In order to evaluate the quantity 
$e_{n}(\{v_{a}^{\mathrm{c}}\},\sigma_{t'}^{2}+N_{0},\mathcal{H})$ 
in (\ref{partition2}), 
we use the fact that $\sigma_{t'}^{2}$ given by (\ref{Sigma_x}) converges 
in probability to $\sigma_{\mathrm{dec}}^{2}(l)\equiv1-\hat{x}_{\mathrm{dec}}^{2}(l)$ 
given by (\ref{squared_decision}) in the limit $K\to\infty$. 
As discussed in \cite{Takeuchi134}, 
the perturbation $\sigma_{t'}^{2}-\sigma_{\mathrm{dec}}^{2}(l)$ provides a negligible 
impact on (\ref{target}) in the large-system limit. Thus, we can replace 
$\sigma_{t'}^{2}$ in (\ref{partition2}) by $\sigma_{\mathrm{dec}}^{2}(l)$ as long as 
the large-system limit is considered. Since 
the soft decisions $\{\hat{x}_{k,T}\}$ are independent zero-mean random 
variables, $\boldsymbol{v}^{\mathrm{c}}=(v_{0}^{\mathrm{c}},\ldots,
v_{n}^{\mathrm{c}})^{\mathrm{T}}$ given $\mathcal{H}$ converges in distribution 
to a CSCG vector with covariance $\mathbb{E}[|\hat{x}_{1,T}|^{2}]\boldsymbol{Q}$, 
given by 
\begin{equation}
\boldsymbol{Q} = \frac{1}{K}\sum_{k=1}^{K}\boldsymbol{h}_{k}(n)
\boldsymbol{h}_{k}(n)^{\mathrm{H}}, 
\end{equation}
where the column vector $\boldsymbol{h}_{k}(n)$ has 
$(\vec{\boldsymbol{h}}_{a})_{k}$ as the $a$th element for $a=0,\ldots,n$.   
It is possible to calculate the conditional expectation over 
$\boldsymbol{v}^{\mathrm{c}}$ after evaluating the integration in (\ref{e}). 
Let us define the function $G(\boldsymbol{Q})$ as 
\begin{equation}
G(\boldsymbol{Q}) 
= -\ln\det(\boldsymbol{I}_{n+1}+\boldsymbol{A}\boldsymbol{Q}) 
- n\ln(\pi N_{0}) - \ln(1+n), 
\end{equation}
with 
\begin{equation}
\boldsymbol{A} = \frac{1}{1+n}
\begin{pmatrix}
n & - \boldsymbol{1}_{n}^{\mathrm{T}} \\ 
- \boldsymbol{1}_{n}^{\mathrm{T}} & (1+n)\boldsymbol{I}_{n} 
-\boldsymbol{1}_{n}\boldsymbol{1}_{n}^{\mathrm{T}}
\end{pmatrix}, 
\end{equation}
where $\boldsymbol{1}_{n}$ denotes the $n$-dimensional column vector whose 
elements are all one. 
Repeating the same argument for the quantity 
$e_{n}(\{v_{a}^{\mathrm{p}}\},N_{0},\mathcal{H})$, we arrive at 
\begin{equation} \label{partition3} 
\frac{1}{K}\ln Z_{t}(n,\omega) 
= \frac{1}{K}\ln\mathbb{E}\left[
 e^{\omega f + K\tilde{G}(\boldsymbol{Q})} 
\right] + O(K^{-1/2}), 
\end{equation} 
where $\tilde{G}(\boldsymbol{Q})$ is given by 
\begin{equation}
\tilde{G}(\boldsymbol{Q}) 
= \frac{T_{\mathrm{tr}}}{K}G\left(
 \frac{\boldsymbol{Q}}{N_{0}} 
\right)
+ \frac{T-T_{\mathrm{tr}}-1}{K}G\left(
 \frac{\hat{x}_{\mathrm{dec}}^{2}(l)}{\sigma_{\mathrm{dec}}^{2}(l)+N_{0}}\boldsymbol{Q} 
\right).  
\end{equation}
In the derivation of (\ref{partition3}), we have used (\ref{squared_decision}). 

The calculation of (\ref{partition3}) was presented in \cite{Takeuchi134} 
under the assumption of replica symmetry (RS). Thus, we omit the remaining 
calculation under the RS assumption.

\ifCLASSOPTIONcaptionsoff
  \newpage
\fi

\balance



\bibliographystyle{IEEEtran}
\bibliography{IEEEabrv,kt-it2014}
\end{document}